\documentclass[showpacs,preprintnumbers,
amsmath,amssymb,aps,prd,nofootinbib,superscriptaddress,
eqsecnum]{revtex4}
\usepackage{amsmath}
\usepackage{graphicx,color}

\makeatletter
\renewcommand{\p@subsection}{}
\makeatother

\renewcommand{\thesection}{\arabic{section}}

\renewcommand{\theequation}{\arabic{section}.\arabic{equation}}

\newcommand{\Slash}[1]{\ooalign{\hfil/\hfil\crcr$#1$}}

\newcommand{\be}{\begin{eqnarray}}
\newcommand{\ee}{\end{eqnarray}}

\begin{document}

\title{Quark Number Fluctuations in a  Chiral Model\\ 
at Finite Baryon Chemical Potential}

\author{C. Sasaki}
\affiliation{%
Gesellschaft f\"ur Schwerionenforschung, GSI,  D-64291 Darmstadt,
Germany}
\author{B. Friman}
\affiliation{%
Gesellschaft f\"ur Schwerionenforschung, GSI,  D-64291 Darmstadt,
Germany}
\author{K. Redlich}
\affiliation{%
Institute of Theoretical Physics, University of Wroclaw, PL--50204
Wroc\l aw, Poland}

\date{\today}

\begin{abstract}
We discuss the net quark and isovector fluctuations as well as
off-diagonal quark flavor susceptibilities along the chiral phase
transition line in the Nambu--Jona-Lasinio (NJL) model. The model
is formulated at non-zero quark and isospin chemical potentials
with non-vanishing vector couplings in the iso-scalar and
iso-vector channels. We study the influence of the quark chemical
potential on the quark flavor susceptibilities in detail and the
dependence of the results on model parameters as well as on the
quark mass. The NJL model findings are compared with recent lattice
results obtained in two--flavor QCD at finite chemical potential.
On a qualitative level, the NJL model provides a consistent
description of the dependence of quark number fluctuations on
temperature and baryon chemical potential. The phase diagram and
the position of the tricritical point in the NJL model are also
discussed for different parameter sets.
\end{abstract}


\setcounter{footnote}{0}

\maketitle


\section{Introduction}
\label{sec:int}

During recent years the phenomenological importance  of
fluctuations in finite temperature/finite density QCD has been
widely recognized \cite{all}. The analysis of fluctuations is a
powerful method for characterizing the thermodynamic properties of
a system. In particular, enhanced fluctuations are an essential
characteristic of phase transitions
\cite{Kunihiro,jb,stephanov,Brown:2001nh,hatta,Fujii:2003bz,probe1}.
Therefore, modifications in the magnitude of fluctuations have been
suggested as a phenomenological probe of deconfinement and chiral
symmetry restoration in heavy-ion collisions \cite{all,probe}. In
this context, fluctuations related to conserved charges, like
baryon number, electric charge and isospin, are particularly
relevant \cite{all}.  This is due mainly to the difference in mass
gap of the effective degrees of freedom in the confined (hadronic)
and deconfined (quark--gluon plasma) phases. Thus, a large change
in charge fluctuations is to be expected when a system is passing
from quark--gluon plasma to hadronic medium \cite{all,probe}. In
heavy ion collisions fluctuations may also reveal information on
expansion dynamics of the medium created in the initial state and
its electromagnetic emissivity as the fluctuations of isospin
charge are connected with space--like screening limit of the
retarded photon self--energy \cite{photon}.

A measure of the intrinsic statistical fluctuations in a system
close to thermal equilibrium is provided by the corresponding
susceptibilities $\chi$. Hence, fluctuations in a thermodynamic
system may be explored, at least in a limited scope, by finding the
dependence of $\chi$ on the thermal parameters. The properties of a
strongly interacting statistical system are characterized by the
temperature and a set of chemical potentials. The latter are
related to the conservation laws, which follow from the global
symmetries of QCD. For an isospin symmetric and electrically
neutral system the thermodynamical ensemble depends only on two
parameters, the temperature $T$ and the quark chemical potential.

One of the consequences of QCD is the existence of a phase
transition between the confined, chirally broken hadronic phase and
the deconfined, chirally symmetric quark-gluon plasma phase. The two
phases are separated by a phase boundary in the $(T,\mu_q)$--plane.
The existence of the phase boundary for $0\leq \mu_q/T\leq 1$ has
recently been established by first principle Lattice Gauge Theory
(LGT) calculations at finite baryon chemical potential
\cite{fodor,lgt1,lgtm,lgtp}.

Arguments based on effective models
\cite{ef1,Klevansky,ef2,ef3,ef4,ef5,ef6,ef7} indicate that at large
$\mu_q$ the transition is first order. On the other hand, for small
$\mu_q$ and two massless quark flavors, the chiral transition is
expected to be second order with the critical exponents of the O(4)
spin model \cite{pisarski}. For finite quark masses, the second
order transition is, due to explicit chiral symmetry breaking, most
probably replaced by a rapid crossover. The different nature of the
phase transition at low and high $\mu_q$ suggests that the QCD
phase diagram should exhibit a critical end point (CEP), where the line of
first order phase transitions end. Beyond the critical end point
the transition would then be continuous, i.e., second order or
crossover. The critical properties of the second order critical end
point are expected to be determined by the Ising model universality
class \cite{ef5,ising}.

The existence of  a critical end point in QCD has been
recently studied in lattice calculations at non-vanishing chemical
potential by either considering the location of Lee--Yang zeros in
(2+1)--flavor QCD  \cite{fodor,shinji} or by analyzing the
convergence radius of the Taylor series for the free energy in
2--flavor QCD \cite{lgt1,lattice:ejiri}. Recent results
\cite{fodor} obtained within the first approach suggest that a
critical endpoint indeed exists and that it is located at $T\simeq
164$ and $\mu_q\simeq 120$ MeV. On the other hand, in ref.
\cite{lattice:ejiri} no direct evidence for the existence of a
critical endpoint has been found in 2--flavor QCD with relatively
large quark masses, in the range where the Taylor expansion is
applicable, i.e., for $\mu_q<T$.

The critical behavior and the position of the critical end point can
also be extracted from observables, that reflect the singular part
of the free energy. Such an observable is the quark susceptibility
$\chi_{ij}$, defined as the second derivative of the
thermodynamical potential $\Omega(T,\vec\mu ,V)$ with respect to
quark--flavor chemical potential $\mu_f$,
\begin{equation}
\chi_{ff^\prime} = - \frac{1}{V}\frac{\partial^2 \Omega}{\partial\mu_f
\partial\mu_{f^\prime}}\,,
\end{equation}
where for two light (u,d)--quarks, $\vec\mu =(\mu_u,\mu_d)$. In
particular, it was recently argued that the net quark number
susceptibility $\chi_q$ may be used to identify the chiral critical
point in the QCD phase diagram
\cite{hatta,Fujii:2003bz,Fujii:2004jt}. 
The flavor off-diagonal susceptibility was also investigated at finite temperature
in perturbative QCD~\cite{pqcd:sus}.
In two--flavor QCD, for an
isospin symmetric system, $\chi_q$ is given by the sum of the $uu$
and $ud$ susceptibilities: $\chi_q=2(\chi_{uu}+\chi_{ud})$.
According to universality arguments, $\chi_q$ diverges  at the
critical end point in the chiral limit as well as for finite quark
masses, while away from the critical end point, the net quark number
fluctuations are expected to be finite. Consequently, one expects a
non-monotonic structure in $\chi_q$ along the phase boundary if the
QCD phase diagram features a critical end point.

The quark susceptibilities have been computed also in lattice QCD
\cite{lattice:ejiri,lgtold,lgtm,gupta}. Recently results for the
quark number and isovector susceptibilities, or equivalently for
the diagonal and off-diagonal quark flavor susceptibilities have
been obtained on a lattice with two light quark flavors using a
p4--improved staggered fermion action with a quark mass $m_q/T=0.4$
\cite{lgt1,lattice:ejiri}. Results for finite quark chemical
potential were obtained by means of a Taylor series expansion. The
susceptibilities were calculated up to fourth order in the quark
chemical potential \cite{lattice:ejiri,rg2,rg3}.

In general the net quark number susceptibility $\chi_q$ shows a
strong suppression of the corresponding fluctuations in the
confined phase and a strong increase with temperature near the
transition temperature $T_0$. Furthermore, the susceptibility near
the transition temperature $T_0$ shows a strong increase with
increasing quark chemical potential, leading to cusp-like structure
at $\mu_q\simeq T$. Besides, the lattice results confirm the
expectation that the isovector quark susceptibility  $\chi_I$ does
not exhibit a peaked structure near the transition temperature and
shows rather weak dependence on $\mu_q$. This behavior of the
susceptibilities is consistent with the existence of a singularity
of the thermodynamic potential close to $\mu_q=T=T_0$, the chiral
end point. However, the enhancement of $\chi_q$ at finite $\mu_q$
below the transition temperature can also be interpreted in terms
of an enhanced contribution from baryon resonances; a very good
description of the $T$ and $\mu_q$ dependences of the various quark
susceptibilities in the confined phase is provided by the resonance
gas partition function \cite{lattice:ejiri,rg1,rg2,rg3}.

Lattice calculations also show a strong correlation between
fluctuations in different flavor components. This is particularly
clear in the LGT results for the off-diagonal susceptibility
$\chi_{ud}$, which shows a strong increase of $u$--$d$ correlations
near the transition temperature $T_0$ and an abrupt loss of
correlations just above $T_0$ \cite{lattice:ejiri,rg2,rg3}.

In this paper we explore the properties of different quark
susceptibilities in terms of an effective chiral model. Of
particular interest is the  characteristics of the quark number
susceptibilities in different channels along the phase boundary and
in the vicinity of the critical end point. The calculations will be
done in the two-flavor Nambu--Jona-Lasinio (NJL) model~\cite{Nambu}
formulated at finite temperature and chemical potentials for the
baryon number and isospin densities.

The NJL model has been used as a model for exploring
qualitative features of the restoration of chiral symmetry in QCD
\cite{Klevansky,Vogl,Klimt:1990ws,Hatsuda,Buballa}. In the
particular case of SU$_c$(2) color symmetry this model has been
recently argued to provide a realistic description of some of the
lattice results \cite{weise}. However, in the SU$_c$(3) case
it is rather unlikely  that the NJL model can provide a
quantitative understanding of LGT thermodynamics, since it does not
exhibit confinement.
\footnote{ Recently an interesting extension of the  NJL model was
proposed that mimics confinement by including the Polyakov line as
dynamical field which couples to constituent quarks \cite{PNJL}.}
Consequently, the hadronic degrees of freedom, in particular the
baryonic resonances, which provide a quantitative interpretation of
the LGT results in the confined phase \cite{rg2,rg3}, are missing
in the model. Furthermore, there are no gluon degrees of freedom in
the NJL model. In QCD thermodynamics, in the chirally
symmetric/deconfined phase, the gluons play an essential r\^{o}le.
Finally, the model suffers from a strong dependence
on the ultra-violet cut-off.

In the chiral limit, the phase diagram of the NJL model shows a
phase separation line, where the spontaneously broken chiral
symmetry is restored. At small densities the transition is second
order, while for an appropriate choice of the coupling constants,
the transition at large densities is first order. Thus, the model
reproduces the gross structure of the phase diagram expected for
QCD. Consequently, the NJL model, formulated at finite T and
$\vec\mu$, can be used to explore the qualitative behavior of quark
susceptibilities and, more generally, universal features of the
chiral phase transition in the $(T,\mu_q)$--plane.

The net quark number susceptibility has been computed in the NJL
model at finite $T$ some time ago \cite{Kunihiro} and recently also
at finite quark chemical potential $\mu_q$ \cite{Fujii:2003bz}. Our
analysis is going beyond previous studies by extending the NJL
Lagrangian to finite quark and isospin chemical potential as well
as to non-vanishing vector coupling among the constituent quarks.
This allows us to model the net quark number susceptibility
$\chi_q$, the isovector one $\chi_I$ or equivalently the diagonal
and off-diagonal quark flavor susceptibilities $\chi_{ff^\prime}$
and to study their dependence on temperature and chemical potential
as well as on model parameters. On a qualitative level the results
can then be confronted with recent lattice results.

The paper is organized as follows: In section~\ref{sec:NJL}, we
introduce  the  NJL model Lagrangian and its  thermodynamics. In
section~\ref{sec:sus}, we introduce the flavor diagonal and
off-diagonal susceptibilities and calculate  their $T$ and $\mu_q$
as well as model parameters dependences.    We  discuss the
qualitative comparison of  the model results  with the recent
lattice findings. Finally in section~\ref{sec:SD}, we give a brief
summary and discussion of our results.


\setcounter{equation}{0}
\section{The two-flavor NJL model}
\label{sec:NJL} 

For two quark flavors  and  three colors  the
Lagrangian of the  Nambu--Jona-Lasinio (NJL) model reads
\cite{Klevansky,Vogl,Hatsuda,Buballa}:
\begin{align}
{\mathcal L}
&= \bar{\psi}( i\Slash{\partial} -m )\psi
 {}+ G_S \Bigl[ \bigl( \bar{\psi}\psi \bigl)^2 +
     \bigl( \bar{\psi}i\vec{\tau}\gamma_5\psi \bigl)^2  \Bigr]
\nonumber\\
&\qquad
 {}- G_V^{\rm (S)} \bigl( \bar{\psi}\gamma_\mu\psi \bigl)^2
 {}- G_V^{\rm (V)} \Bigl[ \bigl( \bar{\psi}\vec{\tau}\gamma_\mu\psi \bigl)^2
 {}+ \bigl( \bar{\psi}\vec{\tau}\gamma_\mu\gamma_5 \psi \bigl)^2 \Bigr]
 {}+ \bar{\psi}\mu\gamma_0\psi\,,\label{eq1}
\end{align}
where $m = \mbox{diag}(m_u, m_d)$ is  the current quark mass
matrix, $\mu = \mbox{diag} (\mu_u, \mu_d)$ the chemical potential
matrix and $\vec{\tau}$ denotes Pauli matrices. The strength of the
interaction between the constituent quarks in the scalar and vector
channels is parameterized by the dimensionful coupling constants
$G_S$, $G_V^{(S)}$ and $G_V^{(V)}$. We note that while the strength
of the scalar-isoscalar and pseudoscalar-isovector interactions are
equal, due to constraints from chiral symmetry, the
vector-isoscalar and vector-isovector interaction terms are
separately invariant, and hence the corresponding interaction
strengths can be chosen independently.
In the following calculations focused on the quark number susceptibility, 
we choose the axial-vector condensate to be zero since it does not couple 
to the vector current.

The  constraints imposed by the conservation of the net quark
number of different flavors are controlled by the chemical
potential $\mu =\mbox{diag}(\mu_u,\mu_d)$. The chemical potentials
for the total net quark density $n_q$ and the iso-vector quark
density $n_I$  are obtained as linear combinations of $\mu_u$ and
$\mu_d$
\begin{equation}
\mu_q = \frac{1}{2}(\mu_u + \mu_d)\,, \qquad \mu_I =
\frac{1}{2}(\mu_u - \mu_d)\,.
\end{equation}
In terms of $\mu_q$ and $\mu_I$, the last term of the Lagrangian
can be expressed as
\begin{equation}
{\mathcal L}_\mu
 = \bar{\psi}\mu\gamma_0\psi = \mu_q\psi^\dagger\psi +
 \mu_I\psi^\dagger\tau_3\psi\,.
\end{equation}

In addition to the current quark masses and the three coupling
constants introduced above, one additional parameter is required to
complete the model. This is the momentum cut-off ($\Lambda)$, which
regulates the ultraviolet divergencies. In vacuum the values of
$\Lambda$ and $G_S$ are fixed by requiring that the pion decay
constant $f_\pi = 92.4$ MeV and the pion mass $m_\pi = 135$ MeV are
reproduced. Choosing the current quark masses $m_u
\simeq m_d = 5$ MeV one finds for a three-momentum cutoff
$\Lambda = 664.3$ MeV and $G_S \Lambda^2 = 2.06$ \cite{Buballa}.

In the chirally broken phase, the ratio of the coupling constants
of $\omega$ and $\rho$ mesons to nucleons is empirically given by
$g_{\omega NN}/\,g_{\rho NN}\simeq 3$. This value is also
consistent with the naive quark model for the nucleon, where the
corresponding couplings to quarks are identical, i.e., $g_{\omega
QQ}/g_{\rho QQ}=1$. We account for this on a qualitative level by
setting $G_V^{\rm (S)}\simeq 3\,G_V^{\rm (V)}$ in the broken phase
and $G_V^{\rm (S)}=G_V^{\rm (V)}$ in the symmetric phase. Thus, we
consider $G_V^{\rm (V)}/G_V^{\rm (S)}$ in the range from $\frac 13$
to $1$, keeping the vector-isoscalar coupling fixed $G_V^{\rm (S)}
= 0.3\,G_S$.

The thermodynamics of the NJL model (\ref{eq1}) at finite
temperature and non vanishing net quark and isospin chemical
potentials is obtained from the partition function
$Z(T,\mu_q,\mu_I,V)$. In the mean field approximation
\cite{Klevansky} the partition function is obtained from the
effective Lagrangian
\begin{align}
{\mathcal L}
&= \bar{\psi}(i\Slash{\partial} - M + \tilde{\mu}\gamma_0 )\psi
 {}- \frac{1}{4G_S}\mbox{tr}\left((M - m)^2\right)
\nonumber\\
&\qquad
 {}+ \frac{1}{4G_V^{\rm (S)}}(\tilde{\mu}_q - \mu_q)^2 {}+
 \frac{1}{4G_V^{\rm (V)}}(\tilde{\mu}_I - \mu_I)^2\,,\label{eq2.4}
\end{align}
where $M=\mbox{diag}(M_u,M_d)$ and the trace $\mbox{tr}$ is in
flavor space.

In Eq. (\ref{eq2.4}) we have introduced  a dynamical mass  $M$ and
a shifted chemical potential $\tilde{\mu}$ given by
\begin{align}
M
&= m - 2G_S\langle \bar{\psi}\psi \rangle\,,
\\
\tilde{\mu} &= \tilde{\mu}_q + \tilde{\mu}_I\tau_3\,,\label{eq2.5}
\end{align}
where
\begin{eqnarray}
\tilde{\mu}_q
&=&
\mu_q - 2G_V^{\rm (S)}\langle \bar{\psi}\gamma_0\psi \rangle\,,
\nonumber\\
\tilde{\mu}_I &=& \mu_I - 2G_V^{\rm (V)}\langle
\bar{\psi}\tau_3\gamma_0\psi\rangle\,.
\label{muq-muI}\label{eq2.6}
\end{eqnarray}
The resulting thermodynamic potential density\footnote{The
thermodynamic potential is given by $\Omega=\omega V$, where $V$ is
the volume of the system.} is of the following form
\begin{align}
\omega (T,\mu;M,\tilde{\mu})
&= \sum_{f=u,d}\omega_f(T,\mu;M_f,\tilde{\mu}_f)+
\frac{1}{4G_S}\mbox{tr}\left((M - m)^2\right)
\nonumber\\
&-
\frac{1}{4G_V^{\rm (S)}}(\tilde{\mu}_q - \mu_q)^2 {}-
\frac{1}{4G_V^{\rm (V)}}(\tilde{\mu}_I - \mu_I)^2\,,\label{eq2.7}
\end{align}
where
\begin{align}
\omega_f (T,\mu;M_f,\tilde{\mu}_f)&=- 2 N_c
\int\frac{d^3p}{(2\pi)^3}\Bigl[E_f(\vec{p}\,)- T\ln ( 1-n_f^{(+)}(\vec{p},T,\tilde{\mu}_f) )\Bigr.
\nonumber\\
&-\Bigl.  T\ln (1-n_f^{(-)}(\vec{p},T,\tilde{\mu}_f) ) \Bigr]\,.
\label{omega f}
\end{align}

In Eq. (\ref{omega f}), $E_f(\vec{p}\,) = \sqrt{\vec{p}^{\,2} +
M_f^2}$ is the scalar part of the quasiparticle energy. The
contributions of the vector potentials are absorbed in the shifted
chemical potential $\tilde{\mu}_f$. Furthermore, $n_f^{(\pm)}$ is
the distribution function for particle $(+)$ and anti-particle
$(-)$ states
\begin{equation} n_f^{(\pm)}(\vec{p},T,\tilde{\mu}_f) = \Bigl( 1 +
  \exp\bigl[ (E_f(\vec{p}\,) \mp \tilde{\mu}_f)/T \bigr]
  \Bigr)^{-1}\,.
\end{equation}

The condensates appearing in Eqs. (\ref{eq2.5})-(\ref{eq2.7}) are
determined by extremizing the thermodynamic potential\footnote{The
thermodynamic potential is minimized with respect to variations of
the scalar field and maximized with respect to variations of the
(zeroth component) of the vector fields.} with respect to the
dynamical mass and the shifted chemical potentials\ at a given
temperature $T$ and chemical potential $\mu$
\begin{equation}
\frac{\partial\omega}{\partial M}
 = \frac{\partial\omega}{\partial\tilde{\mu}_q} =
 \frac{\partial\omega}{\partial\tilde{\mu}_I} = 0\,.
\label{min cond}
\end{equation}
These conditions yield the scalar condensate and the quark
densities as functions of temperature and chemical potential.
Furthermore, they imply that the scalar and vector fields can be
fixed when computing thermodynamic derivatives. Thus, one
obtains the standard thermodynamic relations for instance for the
quark density
\begin{equation} n_q = -
\frac{\partial\omega(T,\mu)}{\partial\mu}\Big|_T = -
\frac{\partial\omega(T,\mu;M_f,\tilde{\mu}_f)}{\partial\mu}\Big|_{T;M,\tilde{\mu}}\,.
\end{equation}
However, as discussed below, the dependence of $M$ and
$\tilde{\mu}$ on temperature and chemical potential yields
non-trivial contributions to second derivatives of the
thermodynamic potential, e.g., susceptibilities.

The stationarity conditions (\ref{min cond}), together with Eqs.
(\ref{eq2.5})-(\ref{eq2.7}), imply
\begin{align}
M_f
&= m_f + 4G_S N_c \sum_{f=u,d}\int\frac{d^3 p}{(2\pi)^3}
\frac{M_f}{E_f} \Bigl[ 1 - n_f^{(+)}(\vec{p},T,\tilde{\mu}_f) -
n_f^{(-)}(\vec{p},T,\tilde{\mu}_f) \Bigr]\,,
\label{gap eq-mass}
\\
\mu_q
&= \tilde{\mu}_q + 4G_V^{\rm (S)} N_c \sum_{f=u,d} \int\frac{d^3
p}{(2\pi)^3} \Bigl[ n_f^{(+)}(\vec{p},T,\tilde{\mu}_f) -
n_f^{(-)}(\vec{p},T,\tilde{\mu}_f) \Bigr]\,,
\label{muq}
\\
\mu_I
&= \tilde{\mu}_I + 4G_V^{\rm (V)} N_c \int\frac{d^3 p}{(2\pi)^3}
\Bigl[ \Bigl( n_u^{(+)}(\vec{p},T,\tilde{\mu}_u) -
n_u^{(-)}(\vec{p},T,\tilde{\mu}_u) \Bigr) {}- ( u \to d ) \Bigr]\,.
\label{muI}
\end{align}
By comparing Eqs.~(\ref{muq-muI}), (\ref{muq}) and (\ref{muI}), we
find explicit expressions for the quark number density $n_q =
\langle\bar{\psi}\gamma_0\psi\rangle$ and the iso-vector density $n_I = \langle
\bar{\psi}\tau_3\gamma_0\psi \rangle$
\begin{align}
n_q &= 2 N_c \sum_{f=u,d} \int\frac{d^3 p}{(2\pi)^3} \Bigl[
n_f^{(+)}(\vec{p},T,\tilde{\mu}_f) -
n_f^{(-)}(\vec{p},T,\tilde{\mu}_f)
\Bigr]\,,\label{2.16}
\\
n_I &= 2 N_c \int\frac{d^3 p}{(2\pi)^3} \Bigl[ \Bigl(
n_u^{(+)}(\vec{p},T,\tilde{\mu}_u) -
n_u^{(-)}(\vec{p},T,\tilde{\mu}_u)
\Bigr) {}- ( u \to d ) \Bigr]\,.\label{2.17}
\end{align}
In practice, one first solves the gap equation (\ref{gap eq-mass})
for fixed $T$, $\tilde{\mu}_q$ and $\tilde{\mu}_I$ and then
computes $\mu_q$ and $\mu_I$ as well as $n_q$ and $n_I$ using
(\ref{muq})-(\ref{2.17}).

\begin{figure}
\begin{center}
\includegraphics[width=8cm]{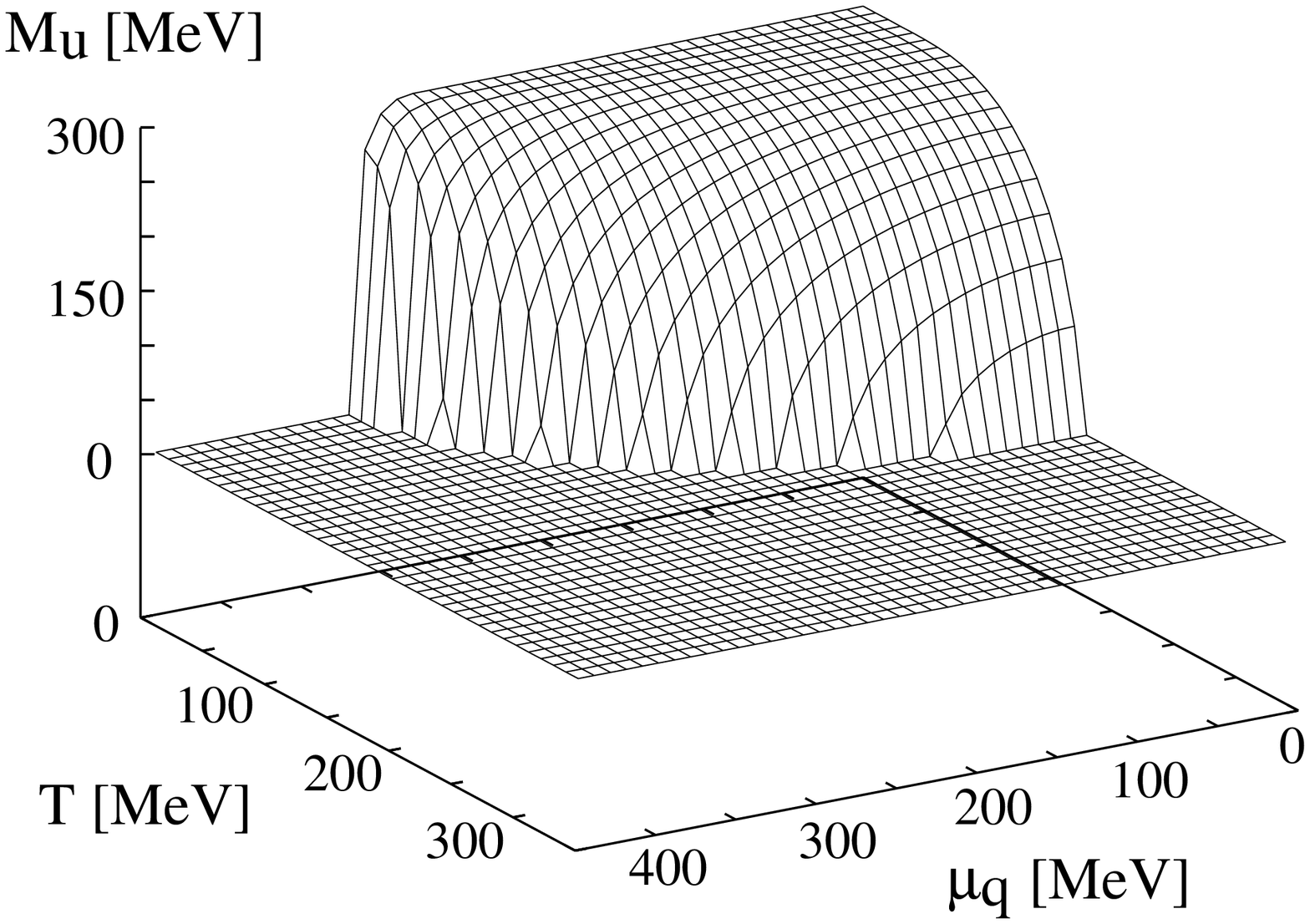}
\includegraphics[width=8cm]{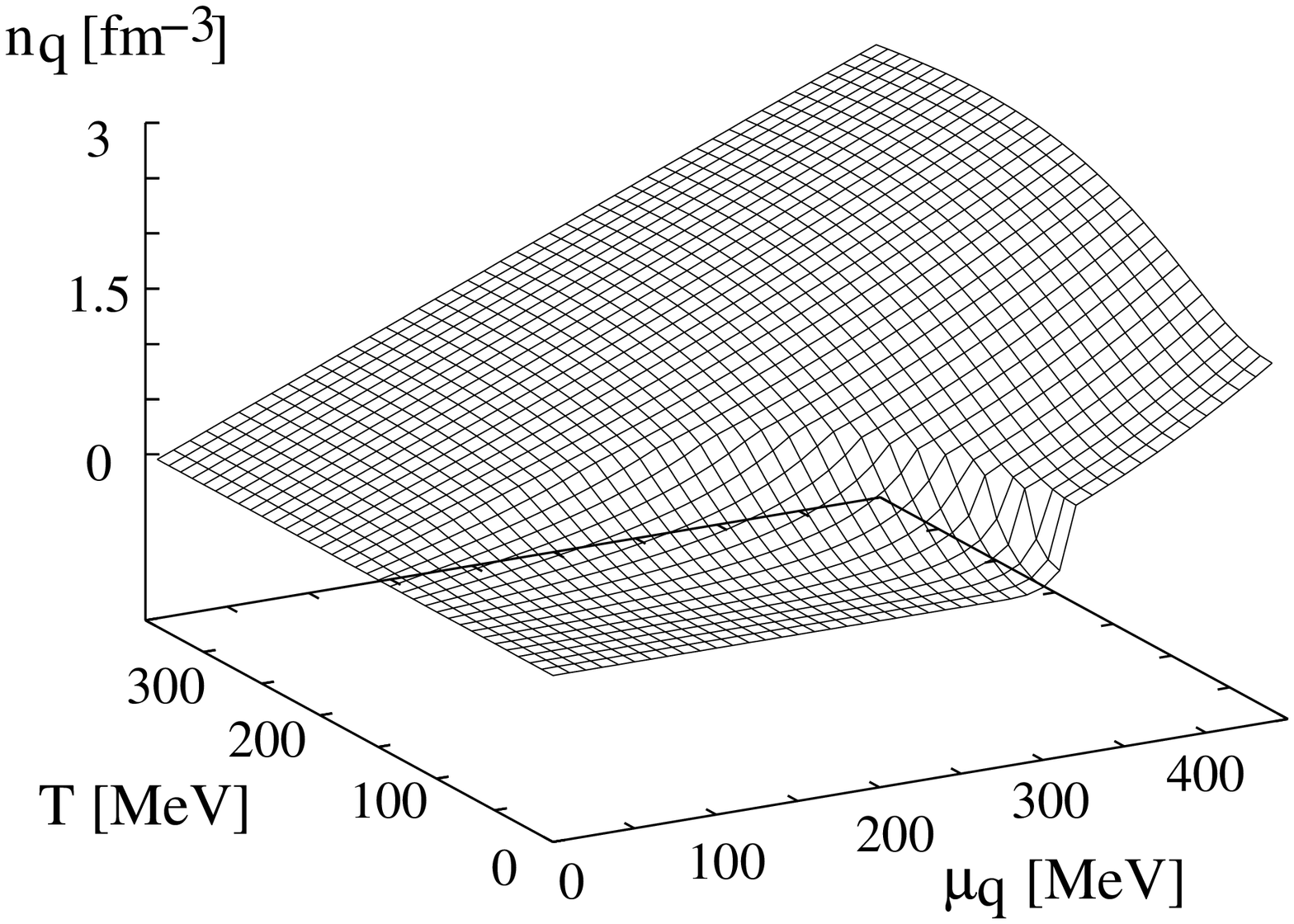}
\caption{\label{fig:Mu} The left-hand figure shows the dynamical
u-quark mass $M_u$ in the chiral limit as a function of temperature
$T$ and the quark chemical potential $\mu_q$. The right-hand figure
represents the quark number density $n_q$ in the chiral limit in
the ($T,\mu_q$)--parameter space. The calculations were done for
isospin symmetric matter, i.e. $\mu_I=0$, with  $G_V^{\rm
(S)}=0.3\,G_S$.
 }
\end{center}
\end{figure}

\begin{figure}
\begin{center}
\includegraphics[width=8cm]{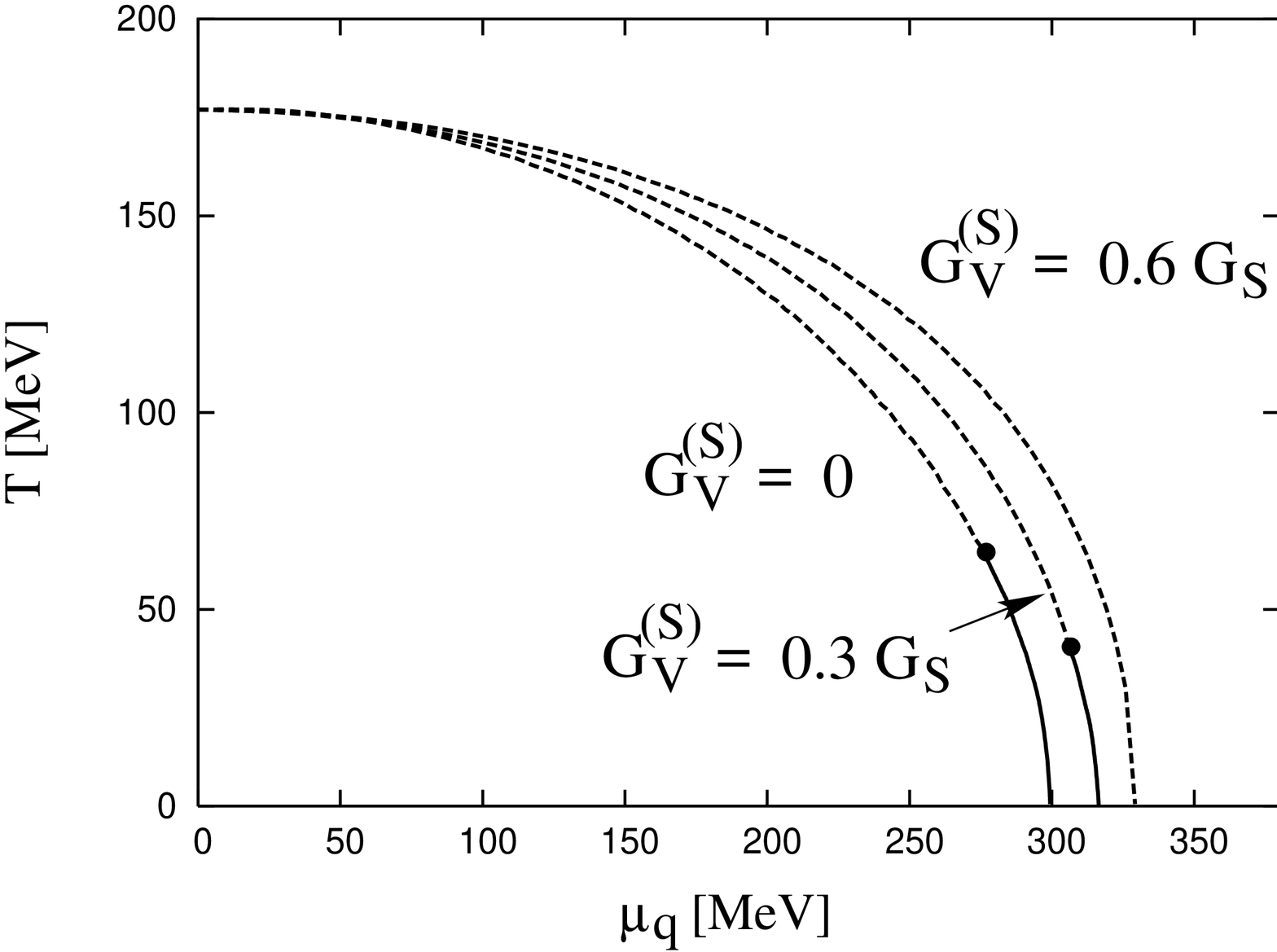}
\includegraphics[width=8cm]{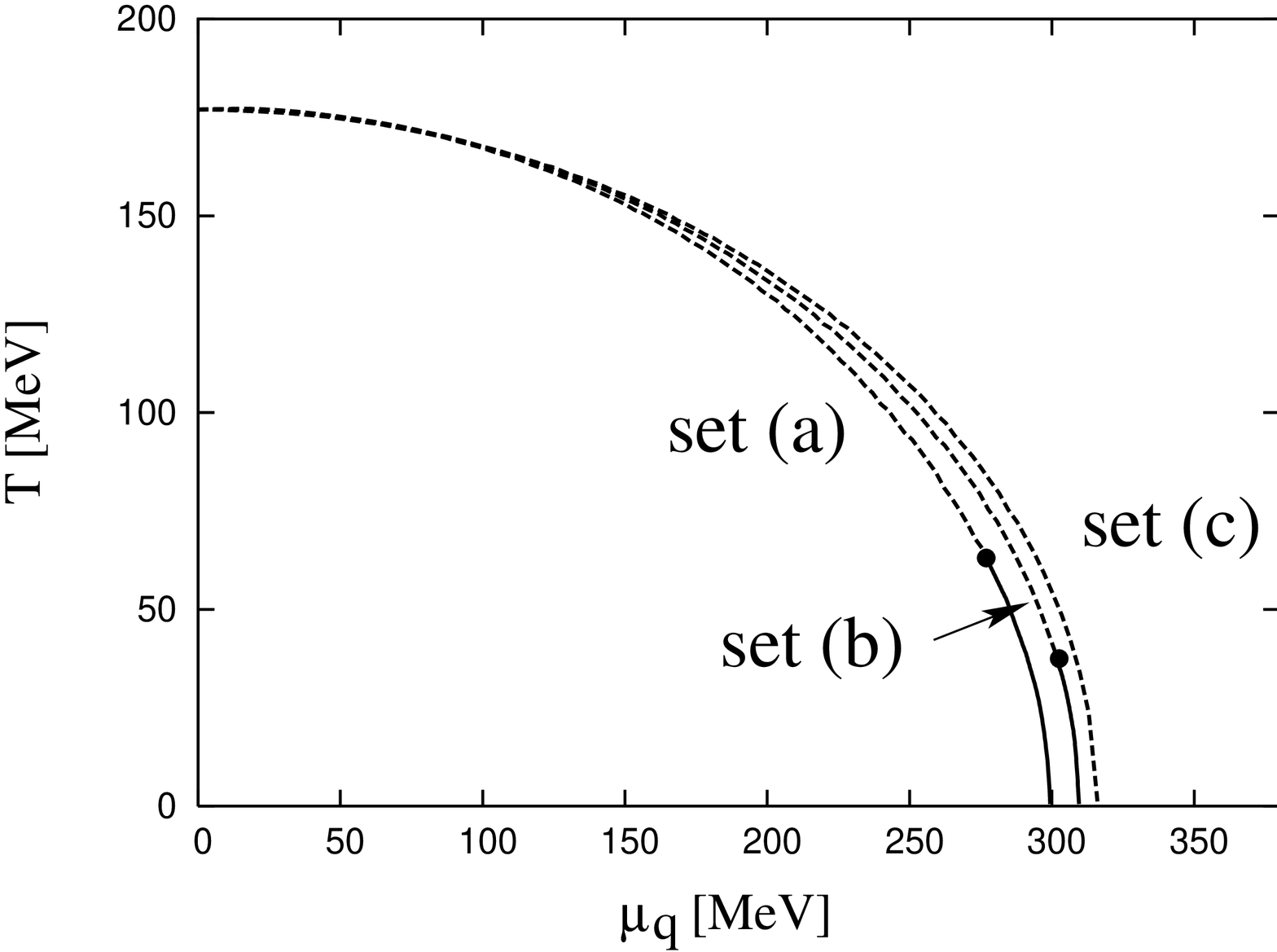}
\caption{\label{fig:phase} The NJL model phase diagram in the
chiral limit for $G_V^{\rm (S)} = 0, 0.3$ and $0.6\,G_S$ (left
panel) and for the set of parameters (a)-(c) of Table. 1 (right
panel). The dashed (solid) line shows the location of the
second-order (first-order) transition. The tricritical point,
indicated by a dot ($\bullet$), is located at $(T,\mu_q)=(65, 275)$
MeV for $G_V^{\rm (S)}=0$ and at $(T,\mu_q)=(42, 305)$ MeV for
$G_V^{\rm (S)}=0.3\,G_S$. Both phase diagrams correspond to a
vanishing isovector chemical potential, $\mu_I=0$.}
\end{center}
\end{figure}

In Fig.~\ref{fig:Mu} we show the  dynamical quark mass $M_f$ and
the net quark number density $n_q$ in the $(T,\mu_q)$--plane for
vanishing iso-vector chemical potential $\mu_I = 0$  in the chiral
limit. For $\mu_I = 0$ the isovector density  $n_I$ vanishes  for
all values of $T$ and $\mu_q$.
In Fig.~\ref{fig:phase} the phase diagram of the NJL model in the
$(T,\mu_q)$--plane is shown for an isospin symmetric system in the
limit of vanishing current quark masses. The boundary between the
chirally broken and symmetric phases was located by finding the
onset of chiral symmetry restoration, $M(T,\mu_q)=0$, when
approaching from the broken phase. As discussed in the
introduction, the order of chiral phase transition is, in the
chiral limit, expected to change from second order at low to first
order at high net baryon densities. Thus, somewhere along the phase
boundary one expects a tricritical point (TCP), where the order of
the chiral transition changes. Close to the phase boundary, the
thermodynamic potential, may be expanded in a power series in the
order parameter $M$ as in Landau theory~\cite{LL}:
$\omega(T,\mu_q;M)=\omega_0+\frac 12 a M^2+\frac 14 b M^4+O(M^6)$.
At a second-order transition $a=0$ and $b>0$, while at a first
order one $a>0$ and $b<0$. (In the latter case, the coefficient of
$M^6$ should be positive for stability.) Thus, the tricritical
point can be identified by $a=b=0$.

In the NJL model the position of the phase boundary and the TCP
depends on the model parameters
\cite{Buballa,Asakawa:1989bq,Lutz:1992dv,Kitazawa:2002bc}. In
Fig.~\ref{fig:phase} we illustrate the dependence on the vector and
scalar couplings $G_V^{\rm (S)}$ and $G_S$ as well as on the
momentum cut-off $\Lambda$. In the left panel the critical line is
shown for different strengths of the vector coupling $G_V^{\rm
(S)}$, keeping $G_S$ and $\Lambda$ fixed. With increasing $G_V^{\rm
(S)}$, the phase boundary at fixed $T$ is shifted to larger
$\mu_q$. This is expected, since at non-zero net baryon density,
the vector coupling $G_V^{\rm (S)}$ provides a repulsive
contribution to the energy of a quark and thus to the chemical
potential (see Eqs.~(\ref{muq-muI}),(\ref{muq})). In fact, in the
$T-\tilde{\mu}_q$ plane the position of the line of second order
phase transitions is independent of $G_V^{\rm (S)}$. This is clear
from the fact that the gap equation Eq. (\ref{gap eq-mass}) depends
on $G_V^{\rm (S)}$ only through $\tilde{\mu}_q$. On the other hand,
the location of the first order transition and the position of the
TCP do depend on the vector coupling $G_V^{\rm (S)}$. This
dependence is reflected in the shift of the TCP to smaller
temperatures with increasing strength of the vector coupling. We
note that for $G_V^{\rm (S)}>0.6\, G_S$, the transition is
everywhere second order and there is no TCP. In the right panel of
Fig.~\ref{fig:phase} we show the dependence of the phase boundary
on the cut off $\Lambda$. We have chosen three parameter sets,
summarized in Table~\ref{table:cutoff}, where the cut off is varied
from our standard value, $\Lambda=664.3$ MeV, up to almost 1 GeV,
keeping $G_V^{\rm (S)}=0$. The corresponding values of $G_S$ are
chosen such that all parameter sets yield the same critical
temperature at $\mu_q=0$, i.e. $T_c = 177$ MeV, in agreement with
lattice results for two--flavor QCD \cite{paikert}. With increasing
$\Lambda$, the phase boundary is again shifted to slightly larger
values of the chemical potential and the TCP to smaller
temperatures.

\begin{table}
 \begin{center}
  \begin{tabular*}{7cm}{@{\extracolsep{\fill}}ccc}
    \hline
    {} &
    $\Lambda$ (MeV) &
    $G_S \Lambda^2$ \\
    \hline
    set (a) &  664.3 & 2.060 \\
    set (b) & 797.2    & 1.935 \\
    set (c) & 995.5    & 1.829 \\
    \hline
  \end{tabular*}
 \end{center}
\caption{Set of cutoff $\Lambda$ and scalar coupling constant
$G_S$ used in the model calculations.} \label{table:cutoff}
\vspace*{0.5cm}
\end{table}

In appendix A we give a closed expression for the integrals
appearing in the gap equation (\ref{gap eq-mass}) at the phase
boundary. Furthermore, by requiring that the critical temperature
at $\mu=0$ remains fixed, as in the right panel of Fig.
\ref{fig:phase}, we obtain a relation between the chemical
potential at the $T=0$ transition and the cut off $\Lambda$
\begin{equation}
\delta\mu=\frac{2}{e^{\Lambda/T}+1}\frac{\Lambda}{\mu}\,\delta\Lambda\, .
\label{mu-shift}
\end{equation}
This relation is derived under the assumption that the transition
is second order everywhere. Nevertheless, it is well satisfied also
when the transition is weakly first order at high quark densities.
Thus, (\ref{mu-shift}) provides a quantitative interpretation of
the shift of the phase boundary found in Fig. \ref{fig:phase}.

We conclude that in the limit of vanishing vector coupling,
$(G_V\to 0)$, the TCP in the NJL model is located at $T\simeq 65$
and $\mu_q\simeq 275$ MeV. For non-zero $G_V^{\rm(S)}$ the position
of the TCP moves towards lower temperature and higher chemical
potential, and for sufficiently large vector coupling, the chiral
transition is second-order everywhere.

The phase diagram in  the $(T,\mu_q)$--plane shown in
Fig.~\ref{fig:phase} applies to the chiral limit. A non-zero quark
mass in the Lagrangian (\ref{eq2.4}) will modify the position of
the phase boundary. Furthermore, a finite quark mass breaks the
chiral symmetry of the Lagrangian explicitly. Consequently, the
second order transition at high $T$ and small $\mu$ is replaced by
a cross--over transition and the TCP by a critical end point. The
critical behavior of the NJL model for finite quark masses is
consistent with the results obtained in other effective models
\cite{hatta,ef2,Asakawa:1989bq,Kitazawa:2002bc,Buballa}.

The position of the phase boundary and the order of the chiral
phase transition can be also identified through thermodynamic
observables. In the Introduction we argued that quark number
fluctuations are sensitive probes of the phase transition.
Furthermore, fluctuations of conserved charges are directly
accessible in experiments. Thus, it is of interest to explore the
behavior of the quark number fluctuations in the vicinity of the
phase boundary in effective models, like the NJL model.

In the next sections we formulate quark susceptibilities in the NJL
model and explore their dependence on thermal and model parameters.
We also consider the influence of finite quark masses on the quark
fluctuations and discuss  the NJL model results in the context of
the  recent lattice findings.


\setcounter{equation}{0}
\section{Quark number susceptibilities}
\label{sec:sus}

The net quark number and iso-vector susceptibilities $\chi_q$ and
$\chi_I$ describe  the  response  of the quark density $n_q$ and
the isovector density $n_I$ to the change of the corresponding
chemical potentials. Thus, $\chi_q$ and $\chi_I$ are defined as
derivatives of $n_q$ and $n_I$  with respect to $\mu_q$ and $\mu_I$
\begin{equation}
\chi_q = \frac{\partial n_q}{\partial\mu_q}\,, \qquad \chi_I =
\frac{\partial n_I}{\partial\mu_I}\,.
\label{eq3.1}
\end{equation}
The net quark and the isovector densities are in the NJL model
given by Eqs. (\ref{2.16}) and (\ref{2.17}). The evaluation of the
derivatives in (\ref{eq3.1}), taking the implicit dependence of the
dynamical masses $M_f$ and the shifted chemical potentials
$\tilde{\mu}_f$ on $\mu_q$ and $\mu_I$ into account, yields
\begin{eqnarray}
\chi_q
&=& \frac{2N_c}{T}\sum_{f=u,d}\int\frac{d^3p}{(2\pi)^3}
    \Biggl[
  {}- \frac{M_f}{E_f}\frac{\partial M_f}{\partial\mu_q}
      \Bigl( n_f^{(+)}\bigl( 1 - n_f^{(+)} \bigr)
        {}-  n_f^{(-)}\bigl( 1 - n_f^{(-)} \bigr) \Bigr)
\nonumber\\
&&\qquad
     {}+ \frac{\partial\tilde{\mu}_f}{\partial\mu_q}
      \Bigl( n_f^{(+)}\bigl( 1 - n_f^{(+)} \bigr)
        {}+  n_f^{(-)}\bigl( 1 - n_f^{(-)} \bigr) \Bigr)
    \Biggr]\,,
\label{sus_q}\label{eq3.2}
\\
\chi_I
&=& \frac{2N_c}{T}\int\frac{d^3p}{(2\pi)^3}
    \Biggl[
  {}- \frac{M_u}{E_u}\frac{\partial M_u}{\partial\mu_I}
      \Bigl( n_u^{(+)}\bigl( 1 - n_u^{(+)} \bigr)
        {}-  n_u^{(-)}\bigl( 1 - n_u^{(-)} \bigr) \Bigr)
\nonumber\\
&&\qquad
     {}+ \frac{\partial\tilde{\mu}_u}{\partial\mu_I}
      \Bigl( n_u^{(+)}\bigl( 1 - n_u^{(+)} \bigr)
        {}+  n_u^{(-)}\bigl( 1 - n_u^{(-)} \bigr) \Bigr)
   {}- (u \to d)
    \Biggr]\,.
\label{sus_I}\label{eq3.3}
\end{eqnarray}

In Eqs. (\ref{eq3.2}) and (\ref{eq3.3}) we have suppressed the $T$
and $\tilde{\mu}_f$ dependence of the distribution functions
$n_f^{(\pm)}$. The derivatives of the dynamical masses
$M_f$ and the reduced chemical potentials $\mu_f$ entering in Eqs.
(\ref{eq3.2}) and (\ref{eq3.3}) are given in
Appendix~\ref{app:der}.

In addition to the fluctuations of the net quark and isovector
densities we also introduce the flavor diagonal and off-diagonal
susceptibilities defined by
\begin{equation}
\chi_{ff}
= -\frac{\partial^2 \omega}{(\partial\mu_f)^2}\,,
\qquad
\chi_{ff^\prime}
= -\frac{\partial^2 \omega}{\partial\mu_f \partial\mu_{f^\prime}}\,,
\end{equation}
with $f \neq f^\prime \in \{ u,d \}$.

In isospin symmetric matter the susceptibilities
$\chi_{uu}(=\chi_{dd})$ and $\chi_{ud}$ are related to $\chi_q$ and
$\chi_I$ by
\begin{equation}
\chi_{uu} = \frac{1}{4}\bigl( \chi_q  + \chi_I \bigr)\,, \qquad
\chi_{ud} = \frac{1}{4}\bigl( \chi_q  - \chi_I
\bigr)\,.\label{eq3.5}
\end{equation}

In the following section we compute the susceptibilities introduced
in Eqs.~(\ref{eq3.2})-(\ref{eq3.5}) in the NJL model and discuss
the dependence of the quark number fluctuations on temperature and
chemical potentials in the vicinity of the phase boundary.


\subsection{Quark susceptibilities in the NJL model}
\label{ssec:NRNJL}

As discussed above, the NJL model does not exhibit the confinement
property of QCD. Thus, there are  no hadronic bound states and
resonances in the chirally broken phase. Instead we are dealing
with constituent quarks which can be viewed as quasi-particles,
with a temperature and density dependent mass. At the chiral
transition the composition of the medium is not changed in the NJL
model; the dynamical quark masses $M_f$ vanish and above $T_c$ the
medium is populated by interacting massless quarks. Furthermore,
high momentum quark modes are suppressed due to the ultra-violet
cut-off. This suppression is particularly relevant at high
temperatures. The differences in the mass spectrum between the NJL
model and QCD, as well as the suppression of high-momentum states,
results in different quantitative properties of the quark number
fluctuations. In spite of these differences, the NJL model is
useful for exploring general features of the susceptibilities close
to the phase boundary and, in particular, near the TCP.

\begin{figure}
\begin{center}
\includegraphics[width=8cm]{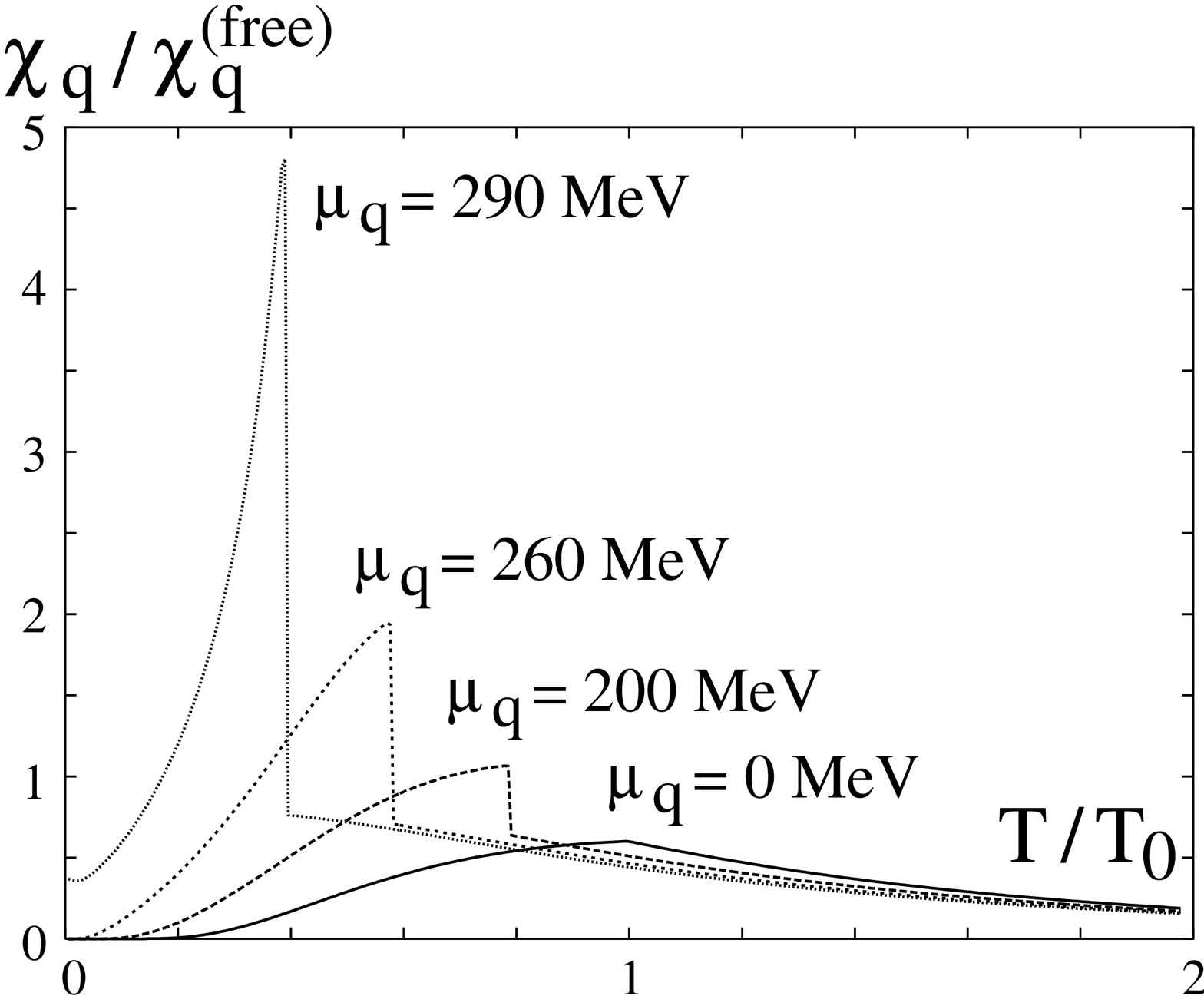}
\includegraphics[width=8cm]{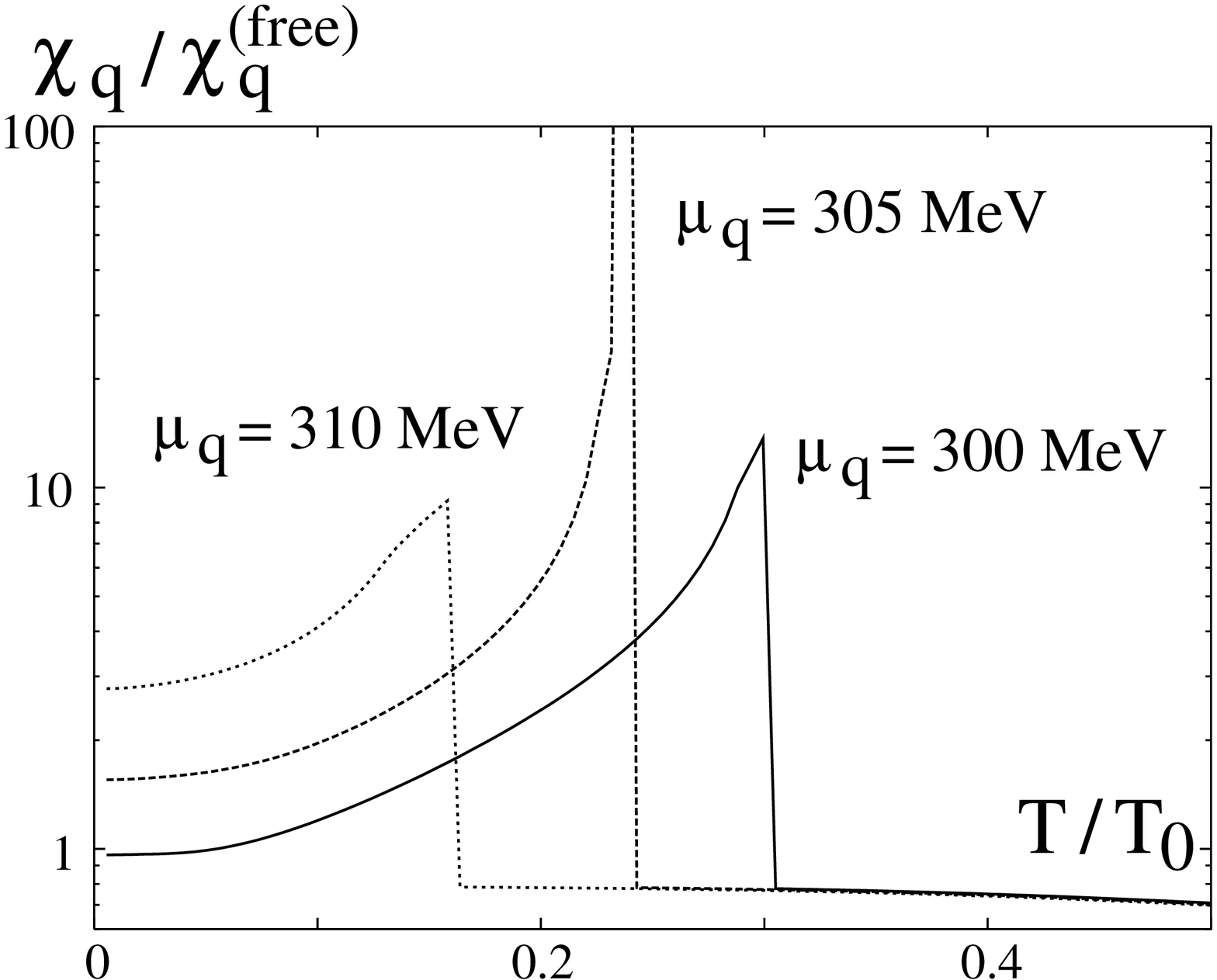}
\end{center}
\caption{\protect\label{fig:chiq} The quark number
susceptibility $\chi_{q}$ as a function of $T/T_0$ in the chiral
limit, for different values of the quark chemical potential
$\mu_q$. Here $\chi_q^{\rm (free)}$ is the quark number
susceptibility for an ideal quark gas and $T_0 = 177$ MeV is the
transition temperature at $\mu_q = \mu_I = 0$. The calculations
correspond to an isospin symmetric system with $G_V^{\rm
(S)}=0.3\,G_S$. } 
\end{figure}

In Fig.~\ref{fig:chiq} we show the quark number susceptibility
$\chi_q$ as a function of $T$  for different values of $\mu_q$,
normalized to that of an ideal quark gas $\chi_q^{\rm (free)}
= N_c N_f (T^2/3 + \mu_q^2/\pi^2)$.

The temperature dependence of $\chi_q$ shows characteristic
features, which vary rapidly with $\mu_q$. The phase boundary is
signaled by a discontinuity in the susceptibility. The size of the
discontinuity grows with increasing $\mu_q$ up to the TCP, where
the susceptibility diverges. Beyond the TCP the discontinuity is
again finite. On the other hand, at $\mu_q=0$ the discontinuity
vanishes and the susceptibility shows a weaker non-analytic
structure at the transition temperature, corresponding to a
discontinuity in $\partial \chi_q/\partial T$. The critical
properties of $\chi_q$ are consistent with a second order phase
transition belonging to the universality class of $O(4)$ spin model
in three dimensions \cite{hatta,rg2}.
Due to the lack of confinement in the NJL model, the leading
contribution to the thermodynamic potential and to fluctuations of
conserved charges is due to single-quark loops. The model can be
improved by including the interaction of quarks with the Polyakov
loops. In the resulting PNJL model~\cite{PNJL,SFR:PNJL}, confinement is
mimicked in the sense that three-quark states are the leading
thermodynamic modes below the phase boundary, in the ``confined''
phase. In this model the fluctuations of the net quark number in
the chirally broken phase are suppressed~\cite{SFR:PNJL} compared to the
results of the NJL model shown in Fig. 3. Furthermore, in the PNJL
model the dependence of the fluctuations on the value of the quark
chemical potential is also stronger. Recently, it was shown
\cite{recent} that by an appropriate choice of the parameterization
of the effective Polyakov loop potential, it is possible to
quantitatively reproduce some LGT results on fluctuations of
conserved charges. Nevertheless, the critical properties of the net
quark number fluctuations near the phase transition in the NJL and
in PNJL models are similar, since both models belong to the same
universality class.

We now explore the qualitative features of the critical region
within Landau theory~\cite{LL}. As already indicated in
section~\ref{sec:NJL}, we construct an effective thermodynamic
potential, valid in the vicinity of the chiral transition. The
thermodynamical potential $\omega(M,T,\mu_q)$ is expanded in a
power series in the order parameter $M$, the dynamical quark mass,
around $M= 0$:
\begin{equation}
\omega^{}(T,\mu_q,M) \simeq  \omega_0(T,\mu_q) {}+
\frac{1}{2}a(T,\mu_q)M^2 {}+
\frac{1}{4}b(T,\mu_q)M^4{}+O(M^6)\,.\label{eqgl}
\end{equation}
Here we neglect the $M^6$ term for simplicity. Although it is
crucial for the calculation of critical exponents (see
Appendix~\ref{app:crit}), it does not affect the present argument.
We assume that $b\geq 0$, i.e., that we are above or at the TCP,
where the transition is second order. For $a>0$, the effective
potential (\ref{eqgl}) has a minimum at $M=0$, which corresponds to
the symmetric phase, where $\omega(T,\mu;0)=\omega_0(T,\mu)$. On
the other hand, for $a<0$ the minimum is located at $M_0
=\sqrt{-a/b}$ and the system is in the broken symmetry phase, where
\begin{equation}
\omega^{}(T,\mu_q;M_0) = \omega_0(T,\mu_q)
{}-\frac{1}{4}\frac{a^2(T,\mu_q)}{b(T,\mu_q)}\,.\label{eqglm}
\end{equation}

The second-order phase boundary (the $O(4)$ critical line) is
determined by the requirement $a = 0$ and $b\geq 0$. Above the
critical line, in the symmetric phase, $M=0$ and the quark number
susceptibility  $\chi_q$ is given by
\begin{equation}
\chi_q^{\rm (sym)} =
-\frac{\partial^2\omega_0}{\partial\mu_q^2}\,.\label{eqs}
\end{equation}

The coefficient $a(T,\mu_q)$ may be expanded around any point
$(T_c,\mu_c)$ on the $O(4)$ critical line. Close to the critical
line, it is sufficient to keep only the leading terms
\begin{eqnarray}
a(T,\mu_q) \simeq  A (T - T_c) + B (\mu_q -
\mu_c)\,,\label{eqa}
\end{eqnarray}
where the expansion coefficients $A$ and $B$ depend on $T_c$ and
$\mu_c$.

Using Eqs. (\ref{eqglm}) and (\ref{eqa}) obtains the quark
susceptibility in the broken phase
\begin{equation}
\chi_q^{\rm (broken)} = \chi_q^{\rm (sym)}
{}+ \frac{B^2}{2b(T,\mu_q)}\,,\label{eqas}
\end{equation}
where we have dropped terms that vanish on the critical line. The
discontinuity of $\chi_q$ across the $O(4)$ critical line at finite
$\mu_q$ is given by the second term in (\ref{eqas}). At $\mu_q=0$
the coefficient $B$, and thus the discontinuity of $\chi_q$,
vanishes by symmetry. Keeping the next term in the expansion of
$a(T,\mu_q)\simeq A (T-T_c) + B_2
\mu_q^2$, one finds $\chi_q^{\rm (broken)}(\mu_q=0)=\chi_q^{\rm (sym)}(\mu_q=0) +
(T-T_c)AB_2/b$. Thus, at $\mu_q=0$ the susceptibility at $T=T_c$ is
continuous, while its temperature derivative is discontinuous, as
seen in Fig.~\ref{fig:chiq}.

Finally at the TCP both $a(T,\mu_q)$ and $b(T,\mu_q)$ vanish.
Consequently, the susceptibility in the broken phase (\ref{eqas})
diverges at the TCP, in agreement with the results shown in the
right panel of Fig.~\ref{fig:chiq}. With the present choice of
parameters, the TCP is located at $(T_{\rm \,TCP},\mu_{\rm
\,TCP})=(42,305)$ MeV. Beyond the TCP, the phase transition is first
order. There the susceptibility again exhibits a finite
discontinuity at the phase boundary. The susceptibility in the
symmetric phase, $\chi_q^{\rm (sym)}$, is expected to vary smoothly
along the phase boundary.

From the perspective of heavy ion experiments, several
susceptibilities are of interest. In particular, this applies to
fluctuations of conserved charges, which may be directly accessible
in experiment. Furthermore, those susceptibilities that reflect the
critical behavior, may possibly be used to explore the QCD phase
transition experimentally. As we have stressed repeatedly, the
quantitative structure of the phase diagram and the position of the
critical end point are model dependent. Consequently, in detail the
QCD phase diagram most likely differs from that found in the NJL
model. Nevertheless, such a model study can still answer
phenomenologically relevant questions concerning e.g. the size of
the critical region, where the fluctuations are dominated by the
singularity at the conjectured critical end point.

\begin{figure}
\begin{center}
\includegraphics[width=8cm]{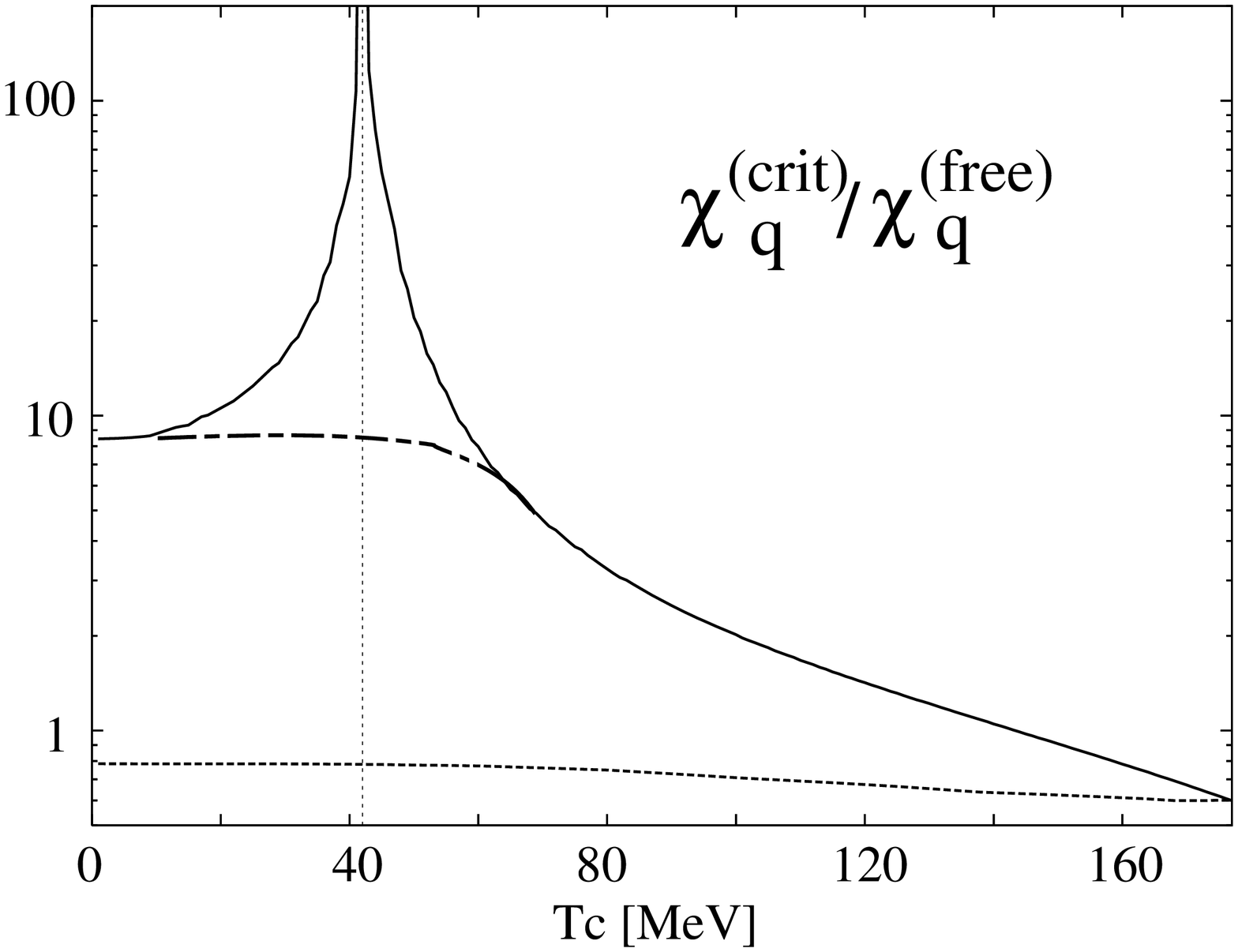}
\includegraphics[width=8cm]{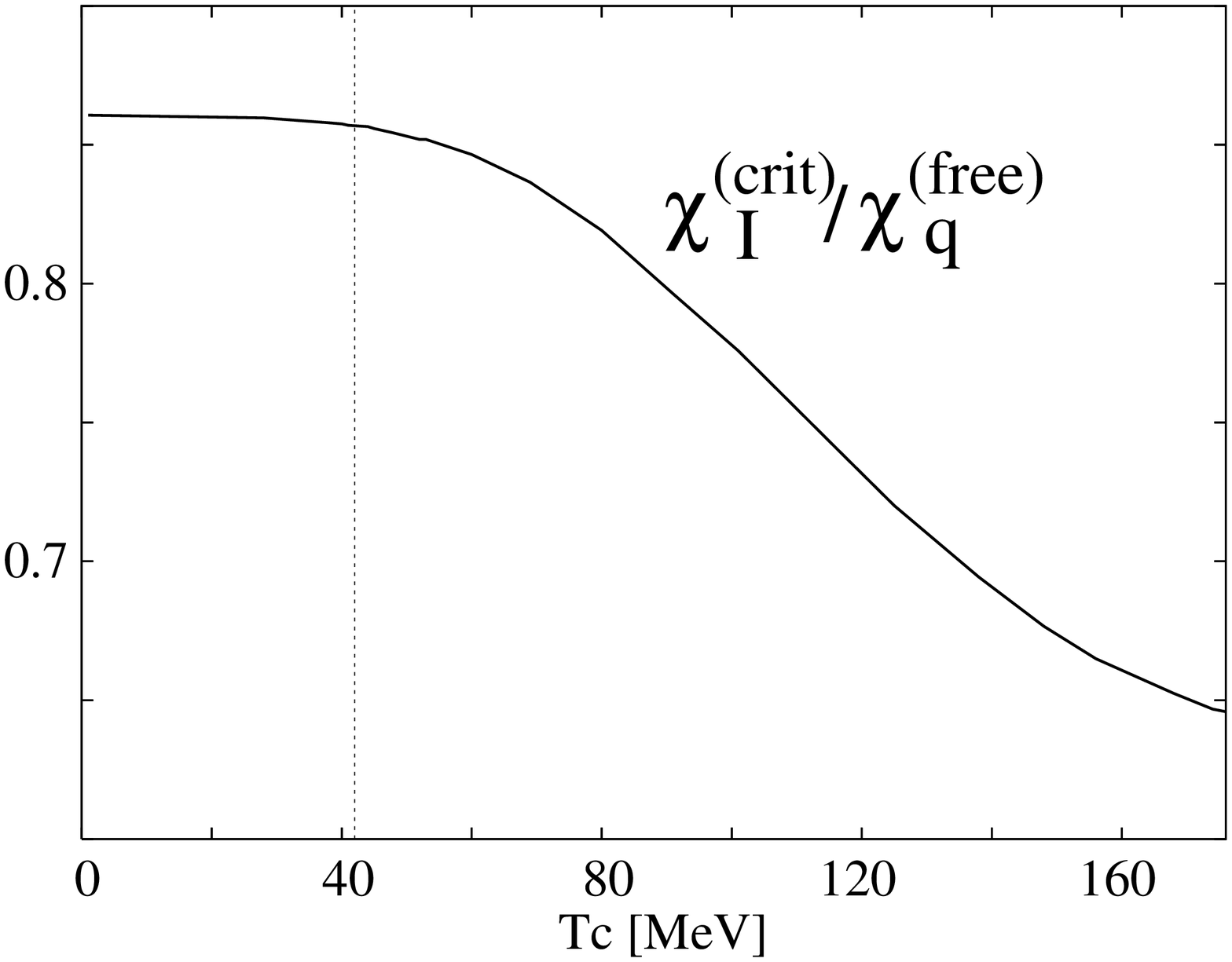}
\end{center}
\caption{\protect\label{tcp_q} The quark number (left) and
isovector (right) susceptibilities $\chi_q$ and $\chi_I$ as
functions of the temperature along the phase boundary. In the
left--hand figure the solid (dashed) line denotes $\chi_q$ in the
chirally broken (symmetric) phase. The vertical dotted-line
indicates the position of the  tricritical point TCP. The
calculations were done in the chiral limit in an isospin symmetric
system with the vector coupling constant $G_V^{\rm (S)}=0.3\,G_S$.
}
\end{figure}

In Fig.~\ref{tcp_q} we show the net quark and isovector
susceptibilities $\chi_q$ and $\chi_I$ along the phase boundary,
given in Fig.~\ref{fig:phase}. The position of the TCP is signaled
by the singularity of the net quark susceptibility $\chi_q$. The
corresponding non-monotonic behavior of the fluctuations 
with increasing beam energy
may give rise to observable effects in heavy-ion
collisions~\cite{stephanov}~\footnote{
 A similar non-monotonic behavior appears in any observable,
 directly related to the net quark number density-density
 correlator. Thus,  measurements of the corresponding non-monotonic
 structure in the baryon number or electric charge  density,
 net-proton number density or in  the mean  transverse momentum
 would be excellent experimental probes of the critical end point in
 the QCD phase diagram.
}. We find that the critical region,
where the fluctuations are dominated by the singularity,
corresponds to a window $\Delta T\simeq 30$ MeV and $\Delta\mu
\simeq 10$ MeV around the TCP. In the absence of a TCP,
the net quark susceptibility would be a monotonic function of $T$
along the phase boundary, as illustrated by the dashed--dotted line
in Fig.~\ref{tcp_q}. We note that the qualitative behavior of the
susceptibility is consistent with the results of Landau theory
discussed above. First, the discontinuity across the phase boundary
vanishes at $\mu_q=0$. Second, the singularity of $\chi_q$ shows up
only in the chirally broken phase, while the susceptibility in the
symmetric phase is monotonous along the phase boundary and shows no
singular behavior.

In nucleus--nucleus collisions a change of the collision energy
$\sqrt s$ is correlated with a corresponding change of the
temperature and the chemical potential. An increase of $\sqrt s$
results in an increase of the temperature $T$ and a decrease of the
baryon chemical potential $\mu_q$. Thus, the critical region around
tricritical point/critical end point $(\Delta T,\Delta \mu)$ can be
(approximately) converted to a range of center-of-mass energies in
A--A collisions. Assuming for simplicity that the relation of $T_c$
and $\sqrt s$ is the same as for the chemical freezeout parameters
extracted from data \cite{our}, we find that $\Delta T\simeq 30$
MeV corresponds to $\Delta \sqrt s\sim 1$ A$\cdot$GeV.
Consequently, this crude estimate implies that in order to observe
effects of critical fluctuations in A--A collisions one would need
to measure an excitation function with an energy step $\Delta\sqrt
s$ smaller than 1 A$\cdot$GeV.

In the calculation of the critical properties we have employed the
mean-field approximation, which in general does not yield the
correct critical exponents \cite{mf}. We note however that the
non-mean-field critical behavior is suppressed near the TCP, since
the quartic coupling $b(T_{TCP},\mu_{TCP})$
vanishes~\cite{hatta,kogut}. Thus, the critical exponents of the
TCP are close to the mean-field exponents~\cite{schaefer-wambach}
and mean-field theory provides a good description of e.g. the
susceptibilities near the TCP. Since the physical quark masses are
small, the critical end point is influenced by the tricritical
point. Thus, one expects the region near the critical end point,
where mean-field theory breaks down, to be relatively
small~\cite{hatta}.

Both at the TCP and at the critical end point the quark-number
susceptibility $\chi_q$ diverges. However, the critical exponents
differ. The mean-field exponents of the TCP and the CEP can be
obtained from Landau theory. As discussed in
Appendix~\ref{app:crit}, the critical exponent for paths
approaching the TCP asymptotically tangential to the phase
boundary, the susceptibility diverges with the critical exponent
$\gamma_q=1$. Approaching the TCP along the first-order transition,
the pre-factor is twice as large as along the O(4) critical line.
For other paths the critical exponent is $\gamma_q=\frac 12$. At
the O(4) critical line, the susceptibility remains finite. The
corresponding critical exponent of the O(4) universality class is
$\alpha\simeq -0.2$, while in the NJL model we obtain the
mean-field value for this critical exponent, $\alpha=0$. Finally we
mention that for non-zero quark mass, at the critical endpoint, the
mean-field critical exponent along a path not tangential to the
phase boundary is $2/3$, while along the phase boundary it remains
equal to unity~\cite{hatta}. When fluctuations are included, the
first exponent is renormalized to that of the 3D Ising model
universality class~\cite{schaefer-wambach}, i.e. $\epsilon=0.78$.

\begin{figure}
\begin{center}
\includegraphics[width=8cm]{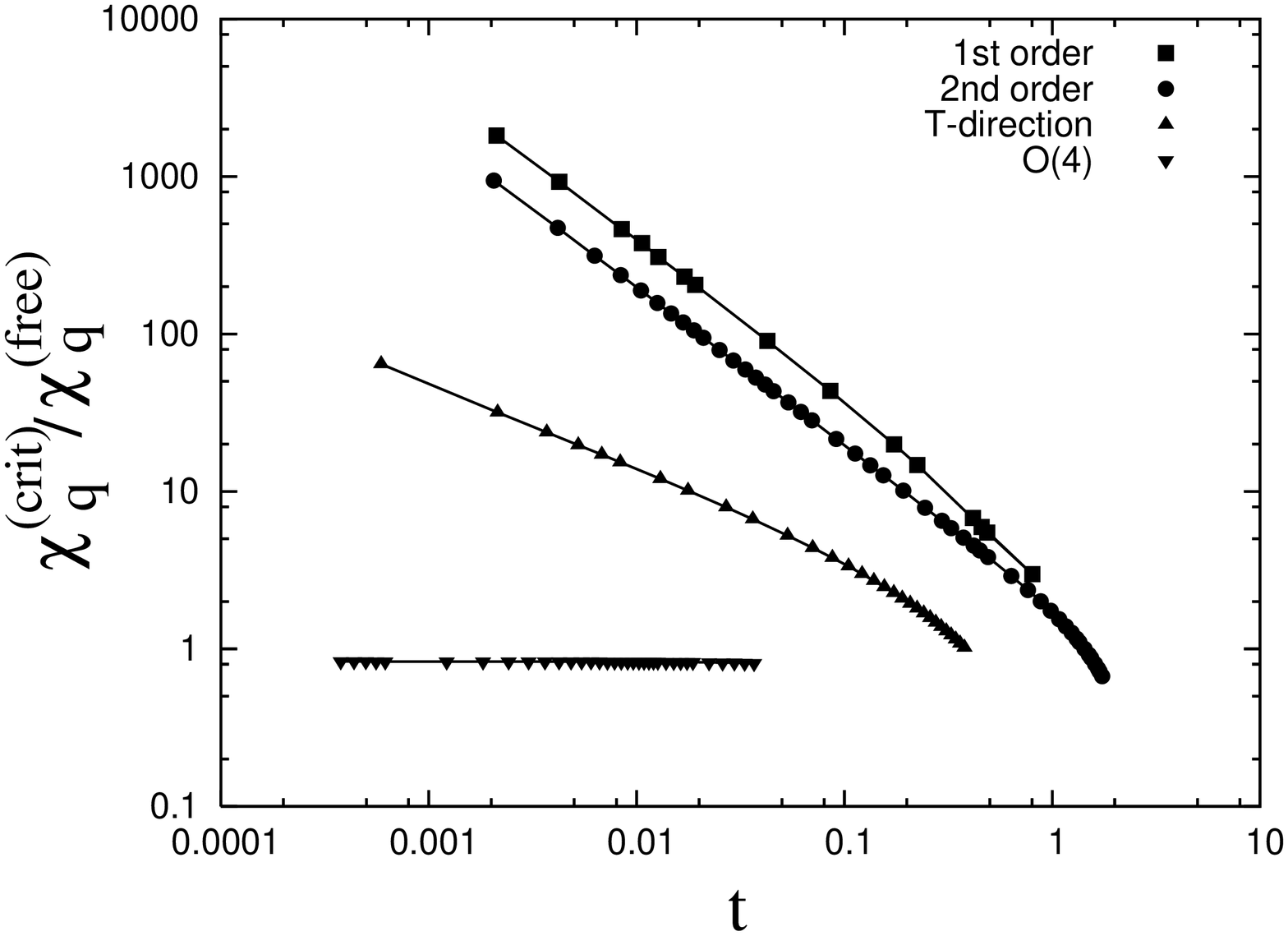}
\includegraphics[width=8cm]{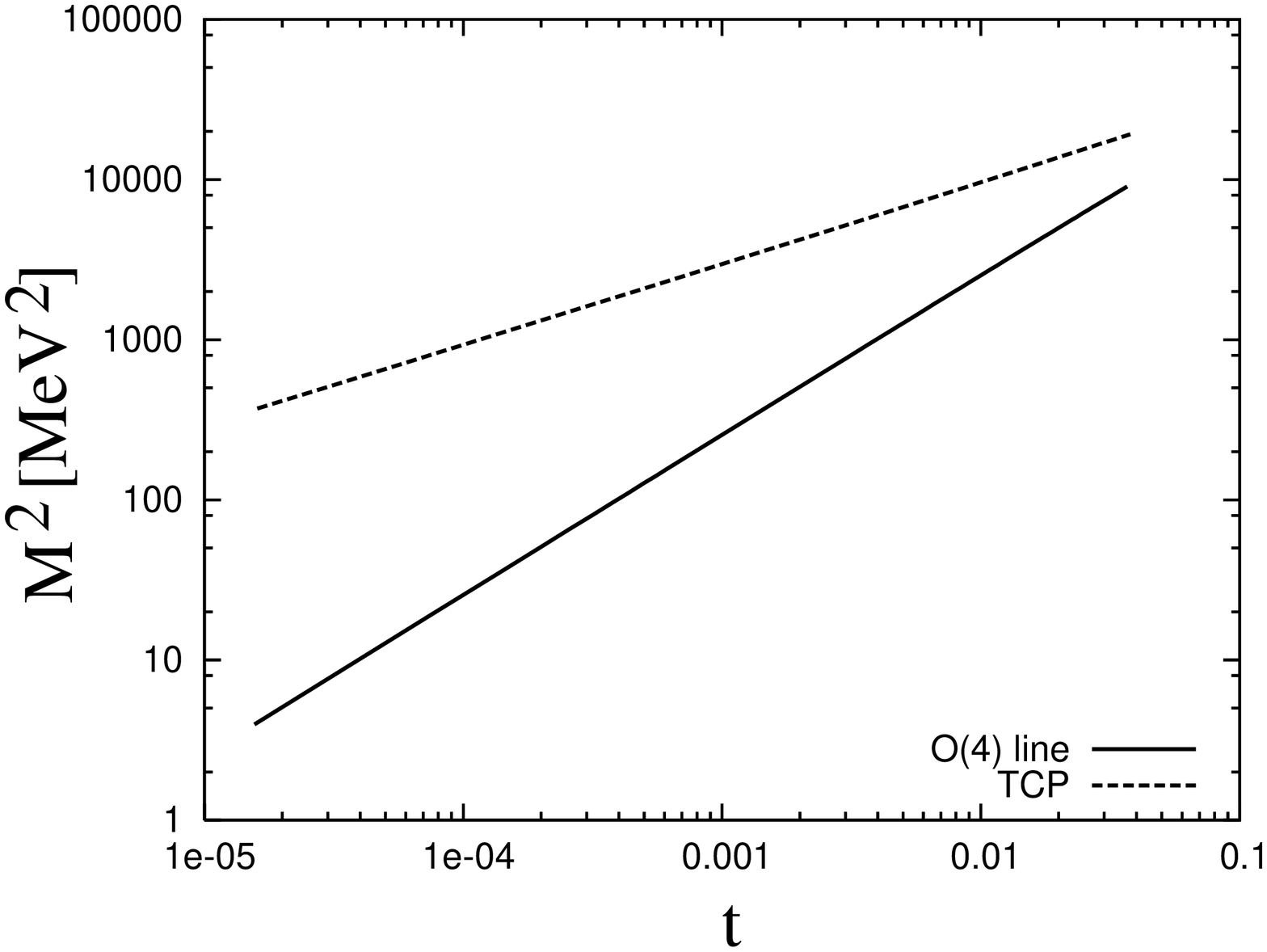}
\end{center}
\caption{\protect\label{crit-exp} The quark number
susceptibility near the tricritical point (left) and the scaling of
the quark mass as the TCP and O(4) critical line is approached
(right). At the TCP the reduced temperature is given by
$t=|T-T_{TCP}|/T_{TCP}$, while at the O(4) line it equals
$t=|T-T_c|/T_c$.}
\end{figure}

In Fig.~\ref{crit-exp} we illustrate the critical behavior near
the O(4) critical line and at the TCP. The dependence on the
reduced temperature $t$ is consistent with the exponents and
relative pre-factors obtained in Landau theory. The different
behavior of the quark number susceptibility at the critical end
point and at the O(4) critical line can be traced to the critical
behavior of the dynamical quark mass. In Appendix \ref{app:crit}
we compute the mean-field exponents for the dynamical quark mass in
Landau theory. At the O(4) critical line $M^2\sim|t|$, while at the
TCP $M^2\sim|t|^{\frac 12}$. The scaling of $M^2$ obtained in the
NJL model are consistent with this, as shown in the right panel of
Fig.~\ref{crit-exp}.

\begin{figure}
\begin{center}
\includegraphics[width=10cm]{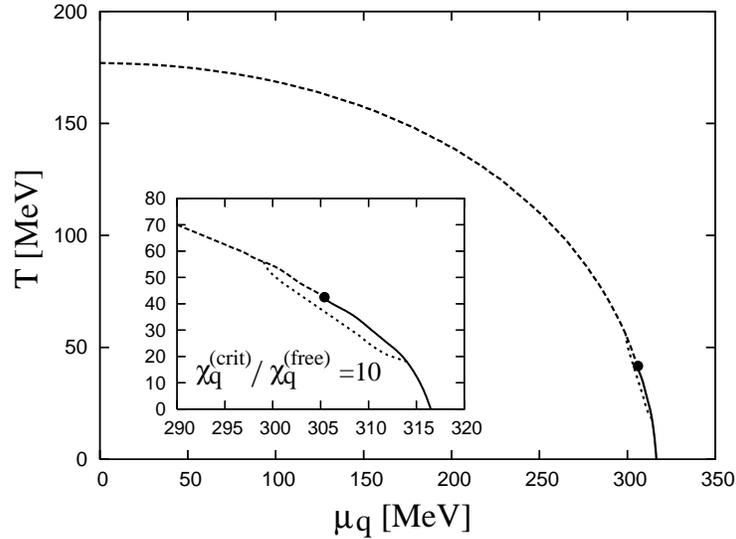}
\end{center}
\caption{\protect\label{crit-reg} The
``critical'' region in the $T-\mu$ plane obtained in the NJL model.
Within this region the susceptibility is enhanced by an order of
magnitude compared to the free one. }
\end{figure}

Furthermore, in Fig.~\ref{crit-reg} we also show the ``critical''
region, where the susceptibility exceeds the free one by more than
an order of magnitude\footnote{By ``critical'' region we mean here
the region where the susceptibility is large due to fluctuations,
not the region of no-mean-field critical behavior.}. The different
critical exponents along the phase boundary and perpendicular to it
are reflected in the shape of the critical region. It is elongated
along the phase boundary, where the singularity is strongest.


In heavy ion collision an additional complication to explore and
map the QCD phase diagram experimentally  appears due to expansion
dynamics, finite system size and secondary hadronic rescattering in
a medium. All these effects can dilute observation of the critical
fluctuations along the chiral phase transition line
\cite{rajagopal}.

\begin{figure}
\begin{center}
\includegraphics[width=10cm]{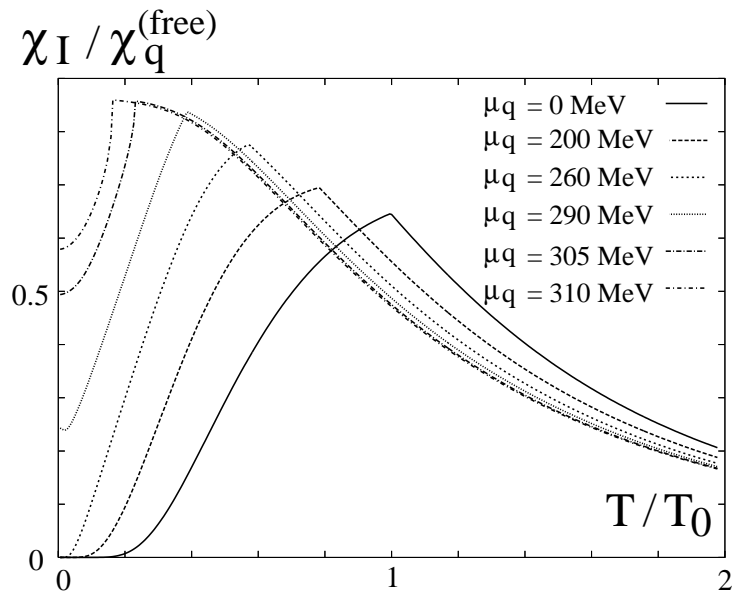}
\end{center}
\caption{\protect\label{fig:chiI}
The isovector susceptibilities
$\chi_{I}$ for different values of net quark chemical potentials
$\mu_q$ as a function of $T/T_0$ in the chiral limit.
The calculations correspond to isospin
symmetric system and the vector coupling constant $G_V^{\rm
(S)}=0.3\,G_S$. } 
\end{figure}

An interesting  observable that characterizes thermal fluctuations
related with isospin conservation is the isovector susceptibility
$\chi_I$ defined in  Eq. (\ref{eq3.1}). The NJL model results for
$\chi_I$ in the isospin symmetric system are shown in
Fig.~\ref{fig:chiI} as a function of $T$ for different $\mu_q$.
The isovector fluctuations, contrary to net quark fluctuations, are
neither singular nor discontinuous at the chiral phase transition
for finite chemical potential. As shown in Fig.~\ref{tcp_q}, we
find a rather smooth increase of $\chi_I$ with increasing $\mu_q$
along phase boundary line. At the TCP where the net quark number
susceptibility diverges, $\chi_I$ remains finite. The non-singular
behavior  of $\chi_I$ at the TCP is consistent with the observation
that there is no mixing between isovector excitations and the
isosclar sigma field due to SU(2)$_V$ isospin
symmetry\cite{stephanovh}. Also recent LGT results
\cite{lattice:ejiri} show a smooth change of the isovector
fluctuations around the deconfinement transition and a fairly weak
dependence of $\chi_I$ on the quark chemical potential $\mu_q$.

The net quark number $\chi_q$ and the isovector $\chi_I$
susceptibilities are related with fluctuations of the electric
charge $\chi_Q$
\be \chi_Q=\frac{1}{36}\chi_q+\frac{1}{4}\chi_I+
\frac{1}{6} {\frac{\partial^2P}{\partial \mu_q\partial\mu_I}}.
\label{eq6} \ee
Here $P$ is the thermodynamic pressure. For isospin symmetric
system the last term in Eq.~(\ref{eq6}) vanishes. Hence in this
case all relevant susceptibilities are linearly dependent. Clearly,
since $\chi_I$ is finite at the TCP, the electric charge
fluctuations $\chi_Q$ diverge with the same critical behavior as
$\chi_q$.
However, at finite $\mu_I$ the properties of $\chi_I$ at the
chiral phase transition in general and at the TCP in particular,
change. At non-vanishing $\mu_I$, the $SU(2)_V$ symmetry is
explicitly broken. Thus, the isoscalar sigma field mixes with the
isospin density \cite{stephanovh}. Consequently, the isovector
susceptibility exhibits a similar structure as $\chi_q$, with a
singularity at the TCP. Furthermore, at finite $\mu_I$ and away
from TCP one expects a peak in $\chi_I$ at the chiral phase
transition. This is due to the exponential dependence of the
susceptibility on $\mu_I$ for massive constituent quarks in the
broken phase and the power-law dependence in the chirally symmetric
phase where the quarks are massless.

\subsection{Model parameter dependence of quark susceptibilities in the chiral limit}

In the previous section, where we considered the net quark and
isovector susceptibilities, the coupling constants of the effective
interaction between constituent quarks were fixed by requiring that
the model reproduces vacuum observables. In the following we
discuss the influence of changes in the model parameters on the
critical properties of the quark flavor fluctuations. We also
present results for the flavor diagonal and off-diagonal
susceptibilities $\chi_{uu}$ and $\chi_{ud}$ and discuss their
properties.

\begin{figure}
\begin{center}
\includegraphics[width=8cm]{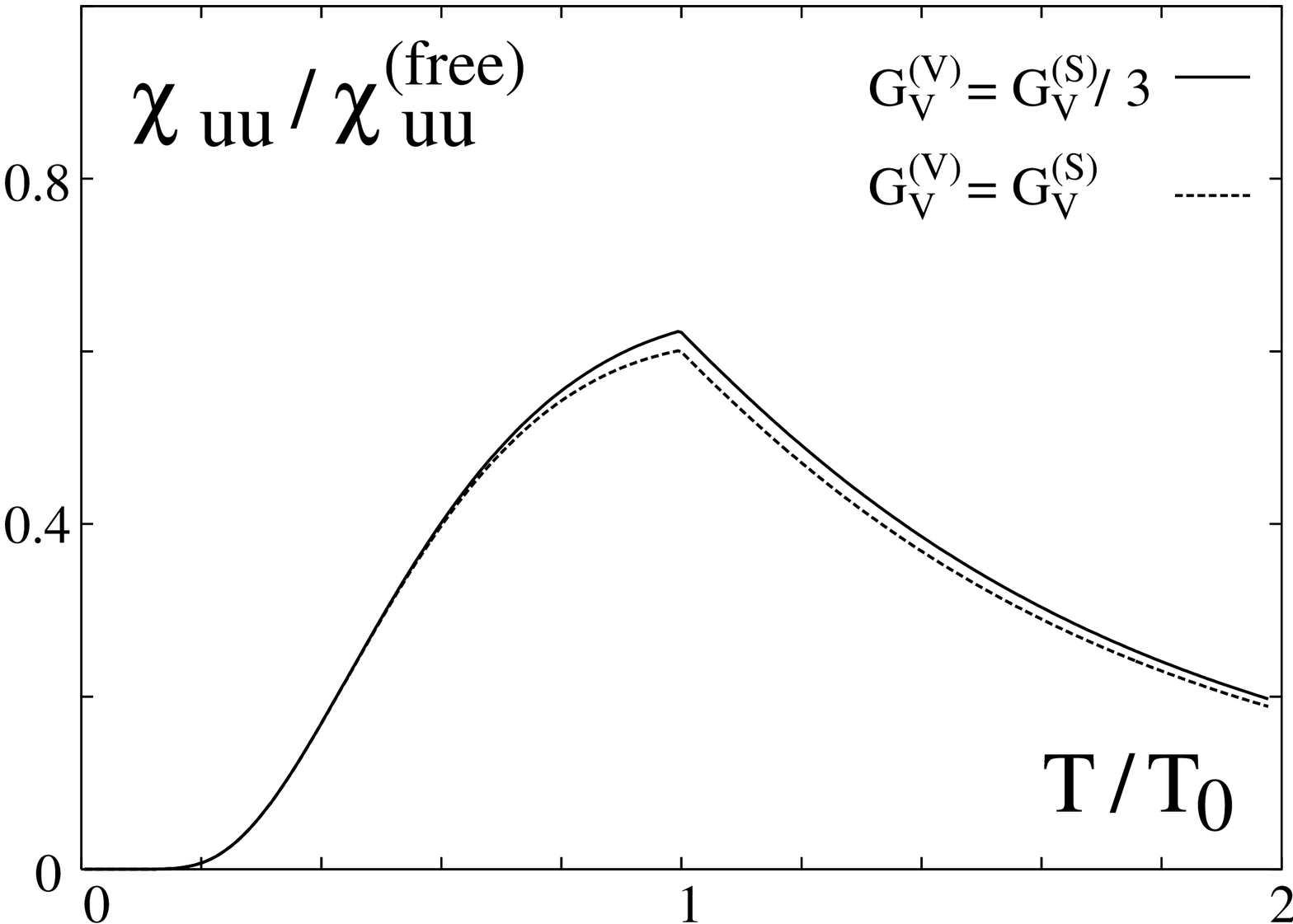}
\includegraphics[width=8cm]{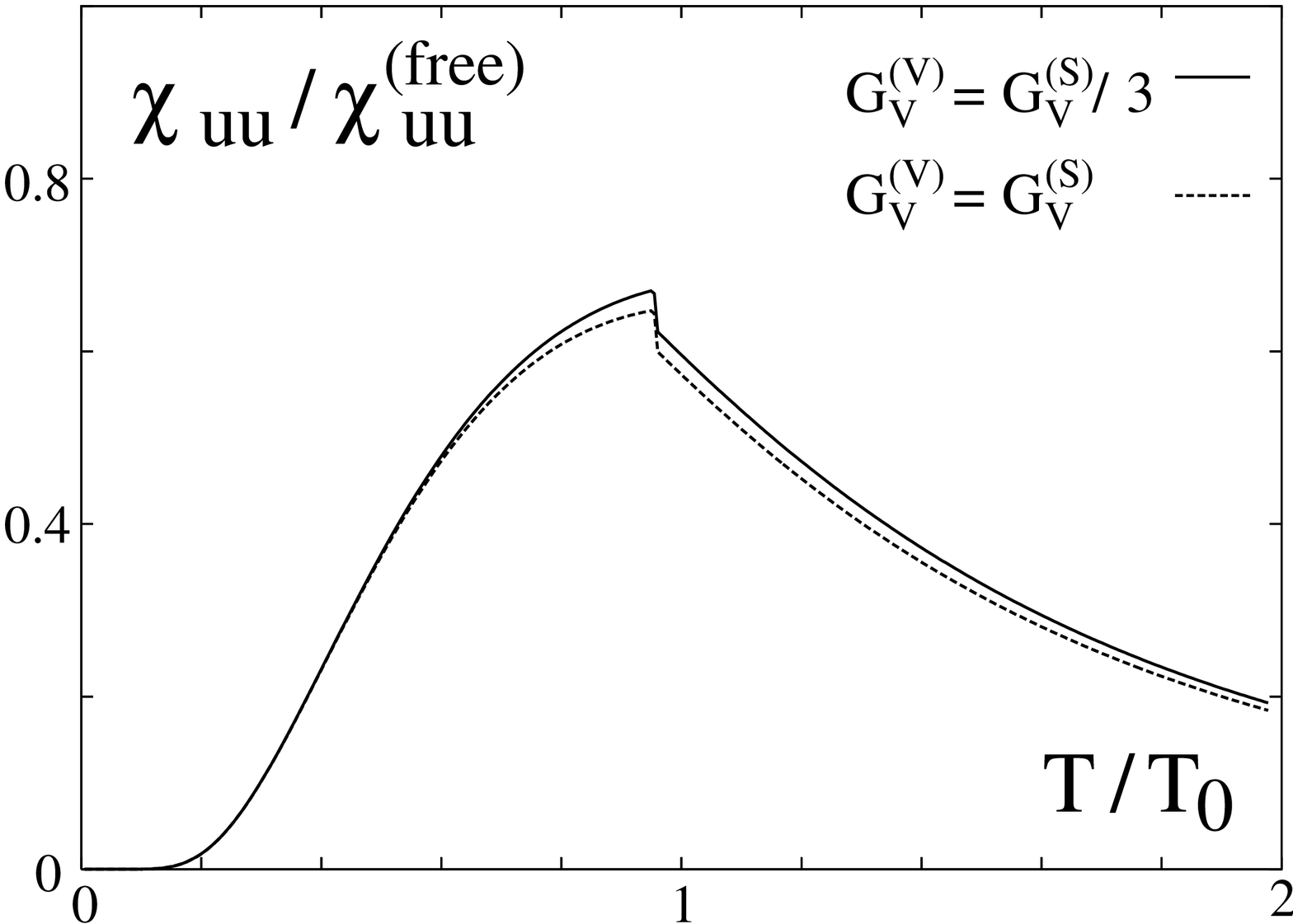}
\\
(a) $\mu_q = 0$ MeV
\hspace*{3.5cm}
(b) $\mu_q = 100$ MeV
\\
\includegraphics[width=8cm]{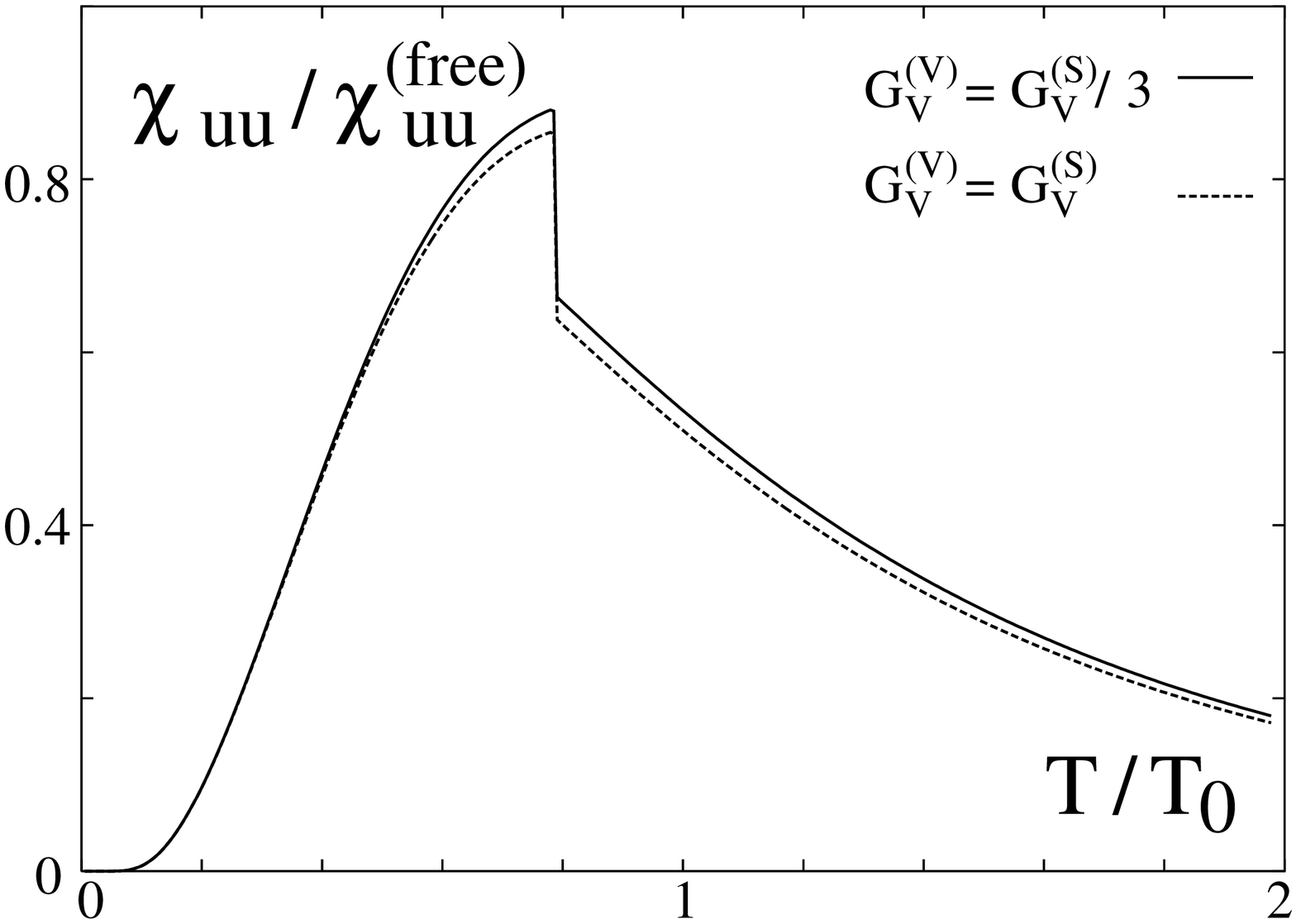}
\includegraphics[width=8cm]{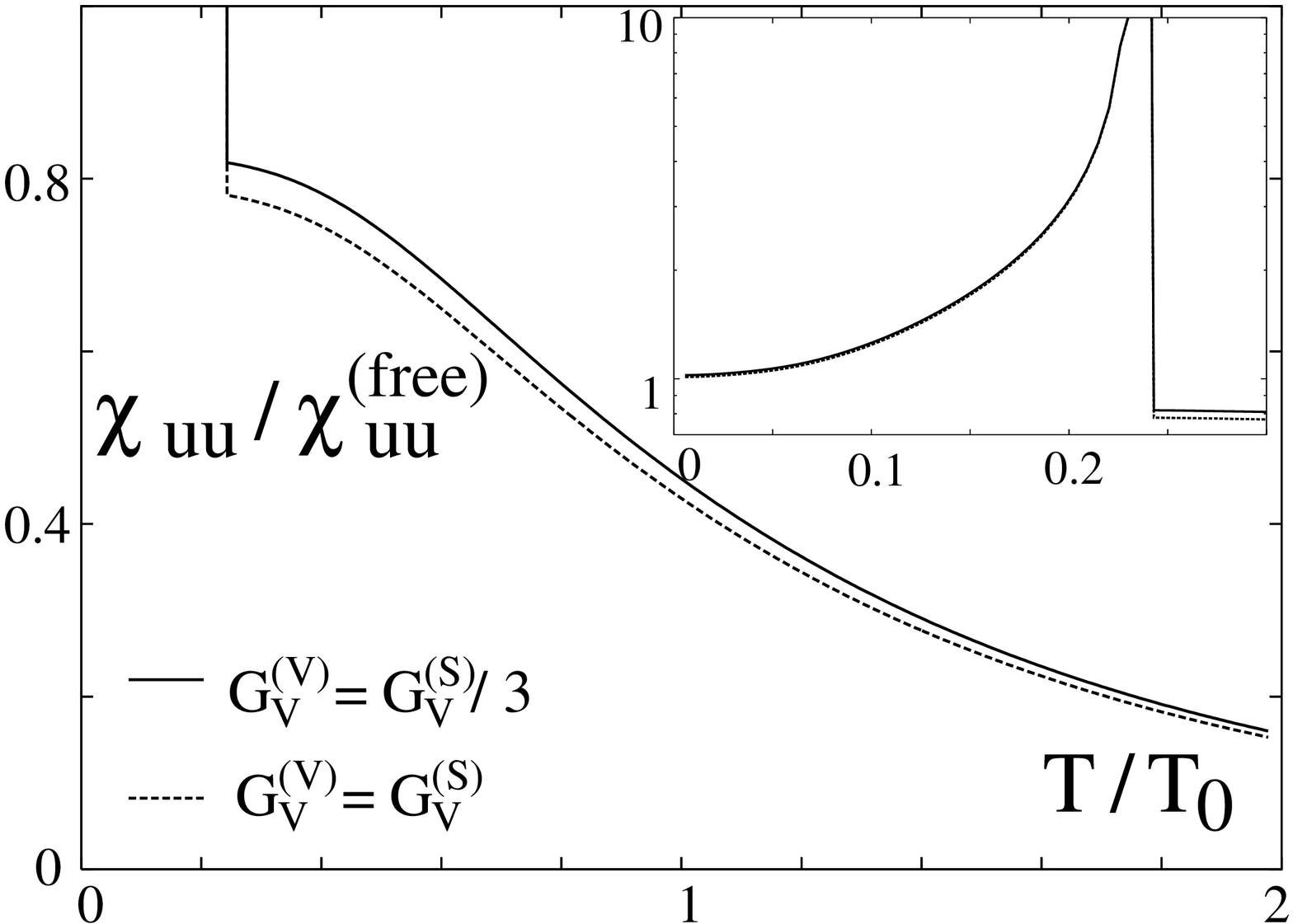}
\\
(c) $\mu_q = 200$ MeV
\hspace*{3.5cm}
(d) $\mu_q = 305$ MeV
\\
\end{center}
\caption{\protect\label{fig:uu-muq} The  diagonal $\chi_{uu}$
susceptibility in an isospin symmetric system for $\mu_q = 0, 100,
200$ and $305$ MeV in the chiral limit normalized to
$\chi_{uu}^{\rm (free)}$ as a function of $T/T_0$. The calculations
correspond to the vector coupling constants $G_V^{\rm
(S)}=0.3\,G_S$ and $G_V^{\rm (V)}/G_V^{\rm (S)}=1/3$ and $1$. }
\end{figure}

In Fig.~\ref{fig:uu-muq} the temperature dependence of the flavor
diagonal susceptibility $\chi_{uu}$ is shown for several values of
the quark chemical potential  for two choices of the vector
couplings $G_V^{\rm (V)}$.  The susceptibility is normalized to the
free one $\chi_{uu}^{\rm (free)}$, which is defined by
$\chi_{uu}^{\rm (free)} = \chi_q^{\rm (free)}/N_f$.

\begin{figure}
\begin{center}
\includegraphics[width=8cm]{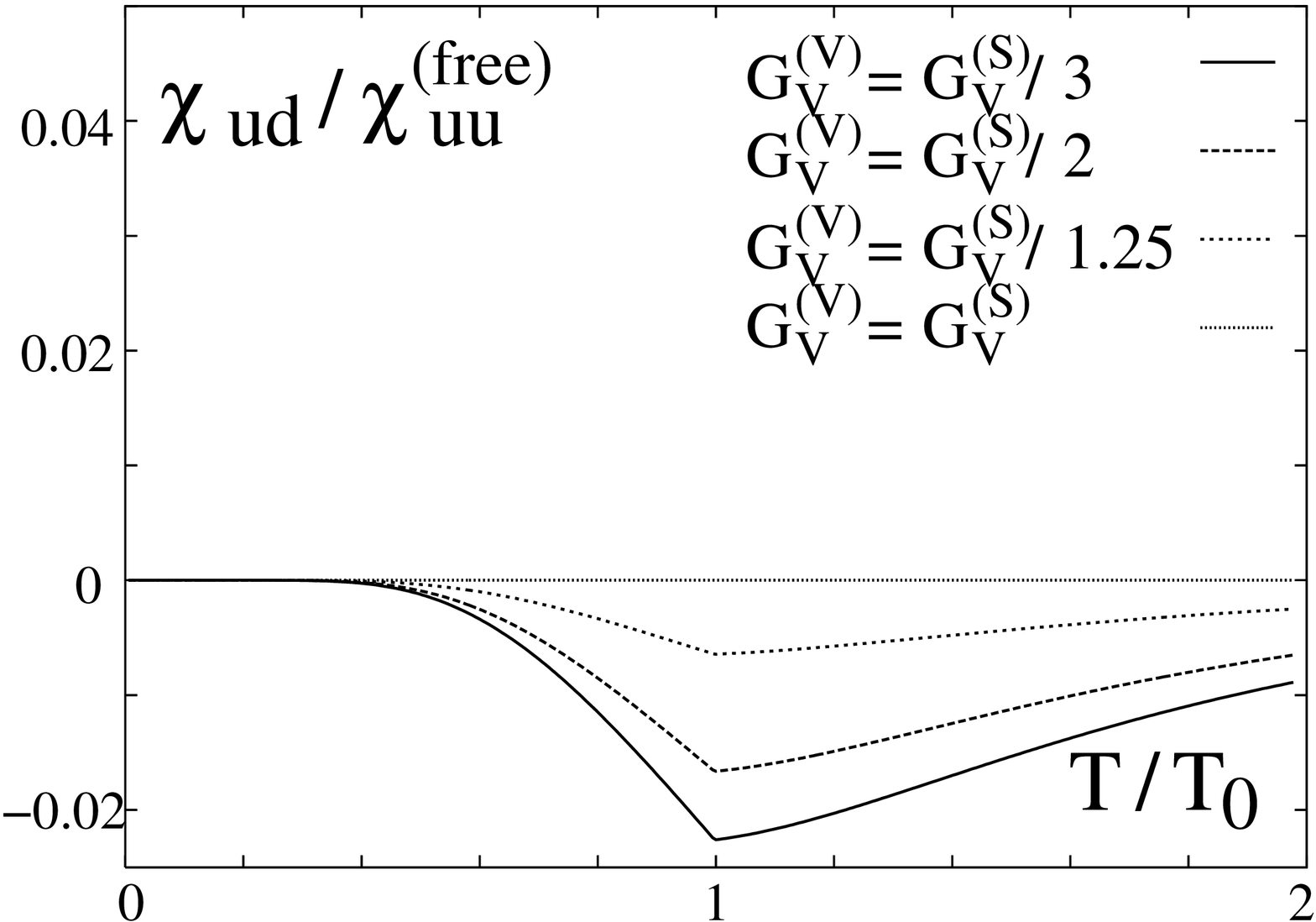}
\includegraphics[width=8cm]{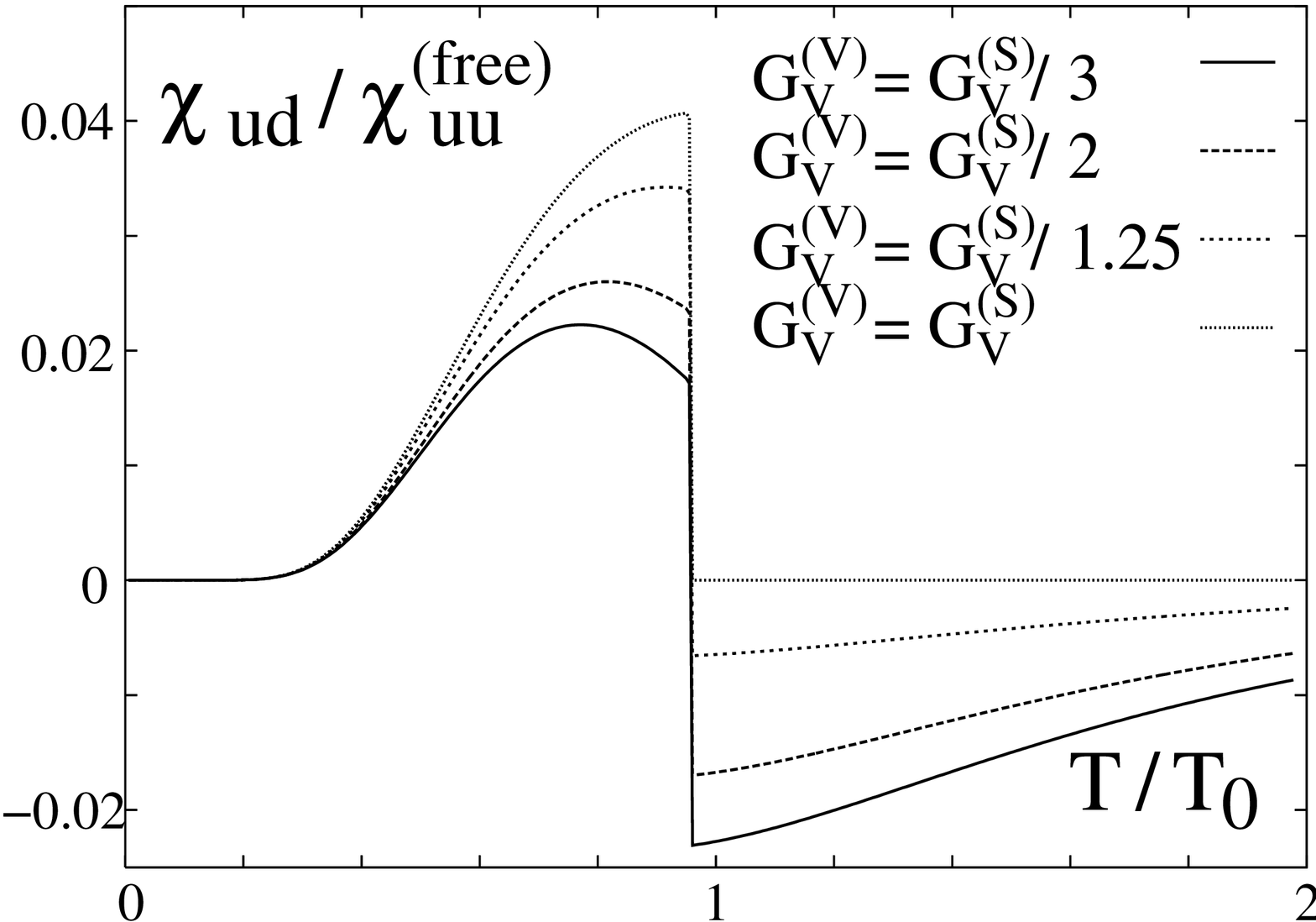}
\\
(a) $\mu_q = 0$ MeV
\hspace*{3.5cm}
(b) $\mu_q = 100$ MeV
\\
\includegraphics[width=8cm]{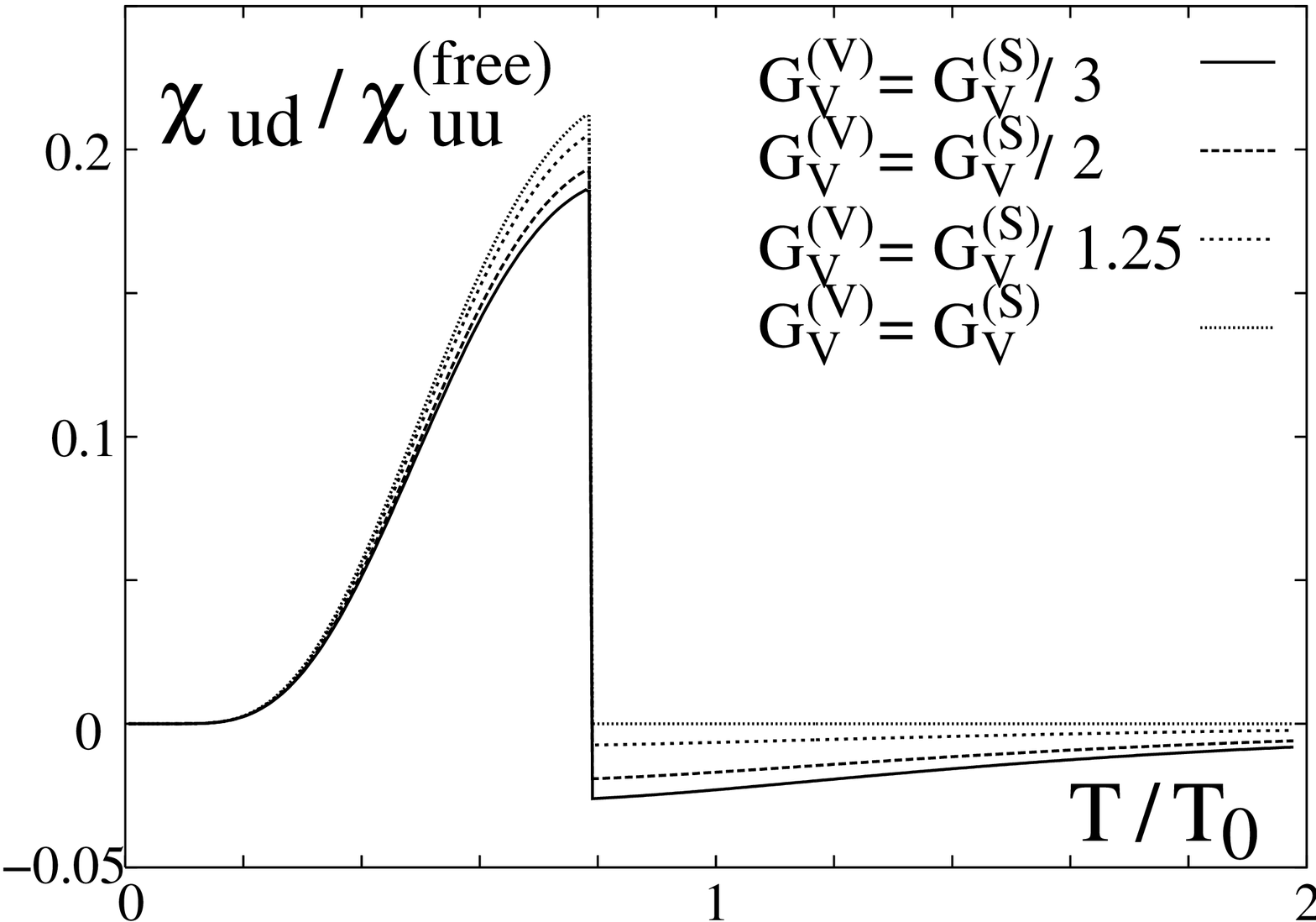}
\includegraphics[width=8cm]{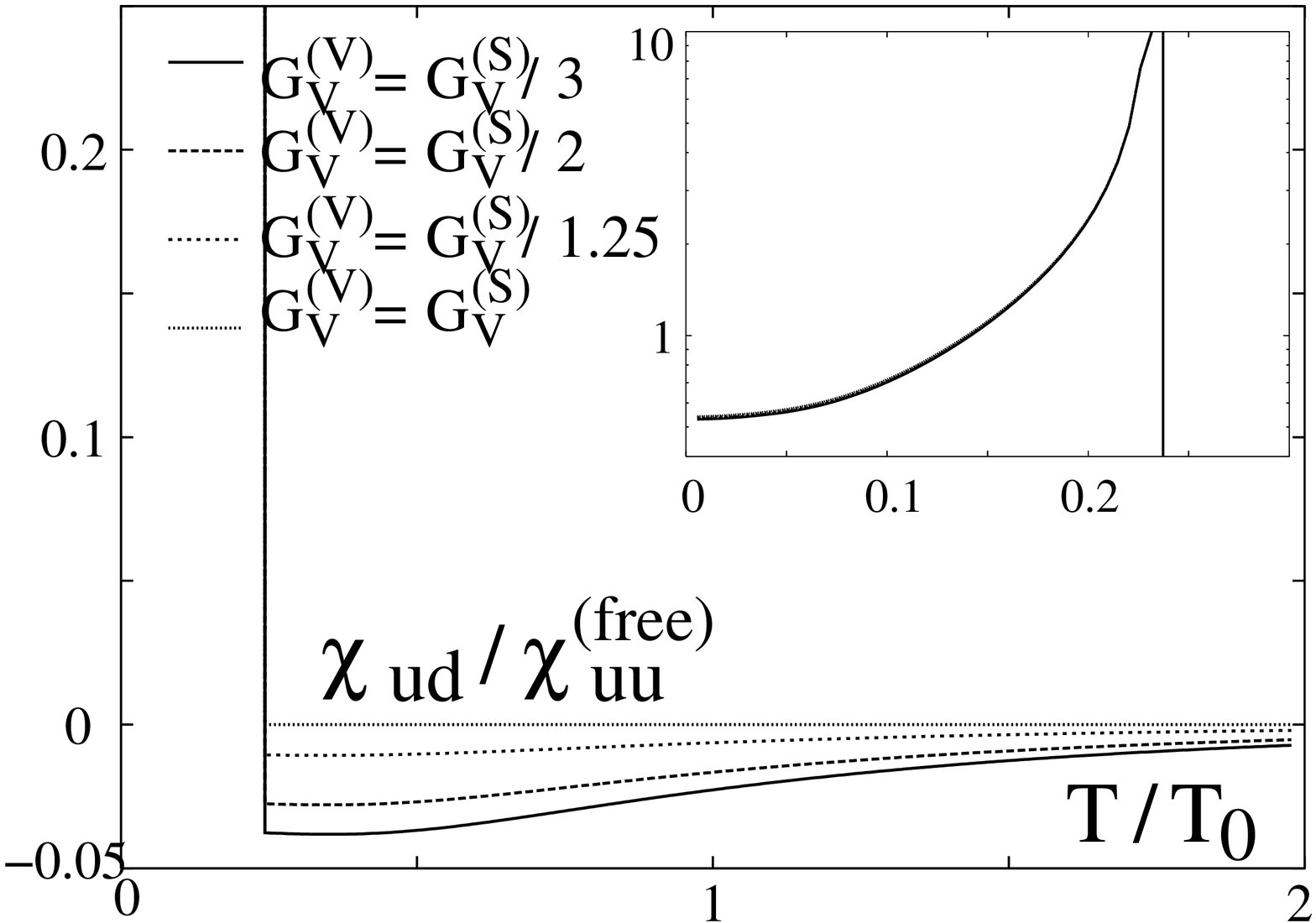}
\\
(c) $\mu_q = 200$ MeV
\hspace*{3.5cm}
(d) $\mu_q = 305$ MeV
\end{center}
\caption{\protect\label{fig:ud-muq} The off-diagonal $\chi_{ud}$
susceptibility in an isospin symmetric system for $\mu_q = 0, 100,
200$ and $305$ MeV in the chiral limit normalized to
$\chi_{uu}^{\rm (free)}$ as a function of $T/T_0$. The calculations
correspond to the vector coupling constants $G_V^{\rm
(S)}=0.3\,G_S$ and $G_V^{\rm (V)}/G_V^{\rm (S)}=1/3$ and $1$.
 }
\end{figure}

\begin{figure}
\begin{center}
\includegraphics[width=8cm]{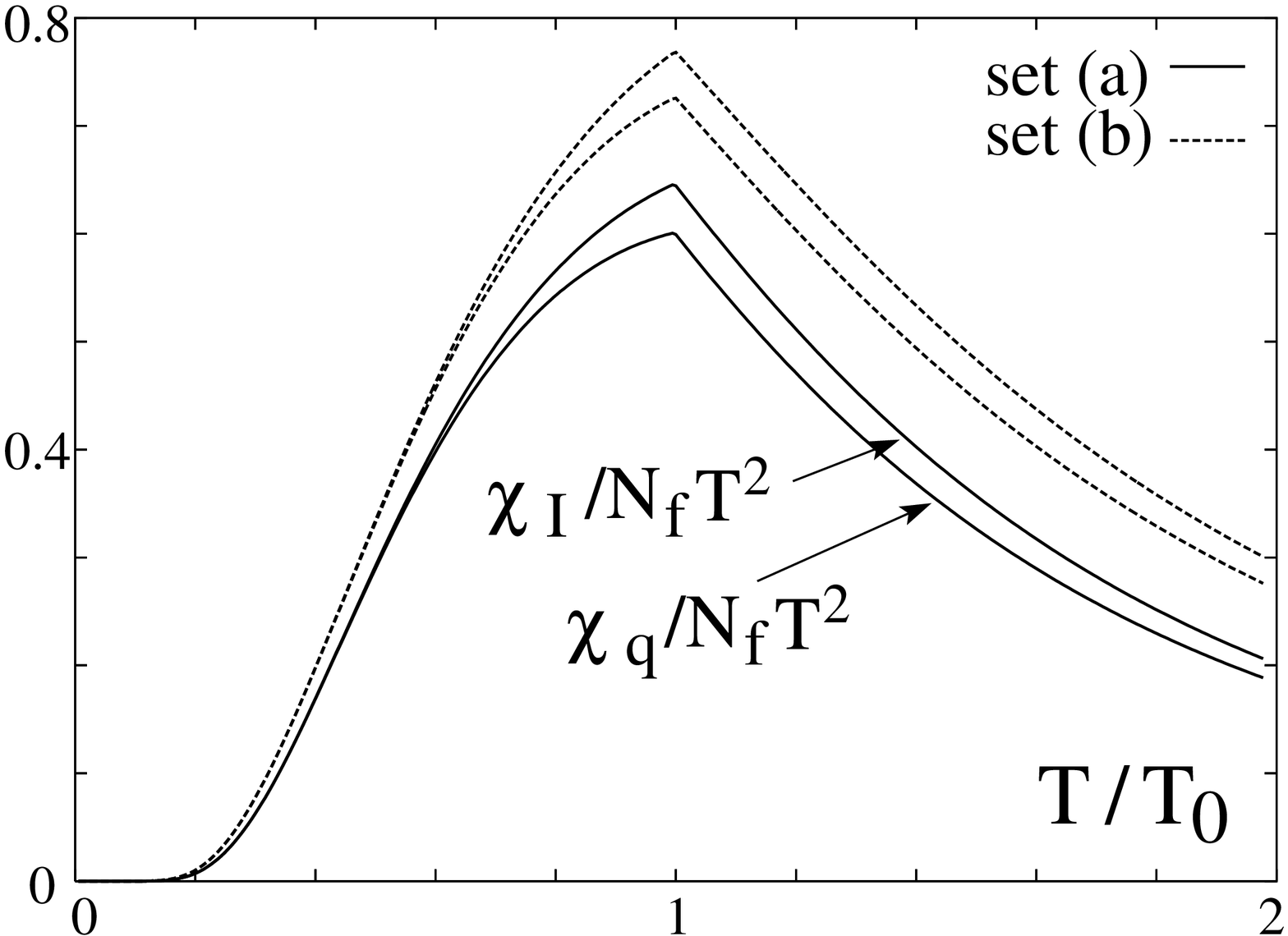}
\\
\includegraphics[width=8cm]{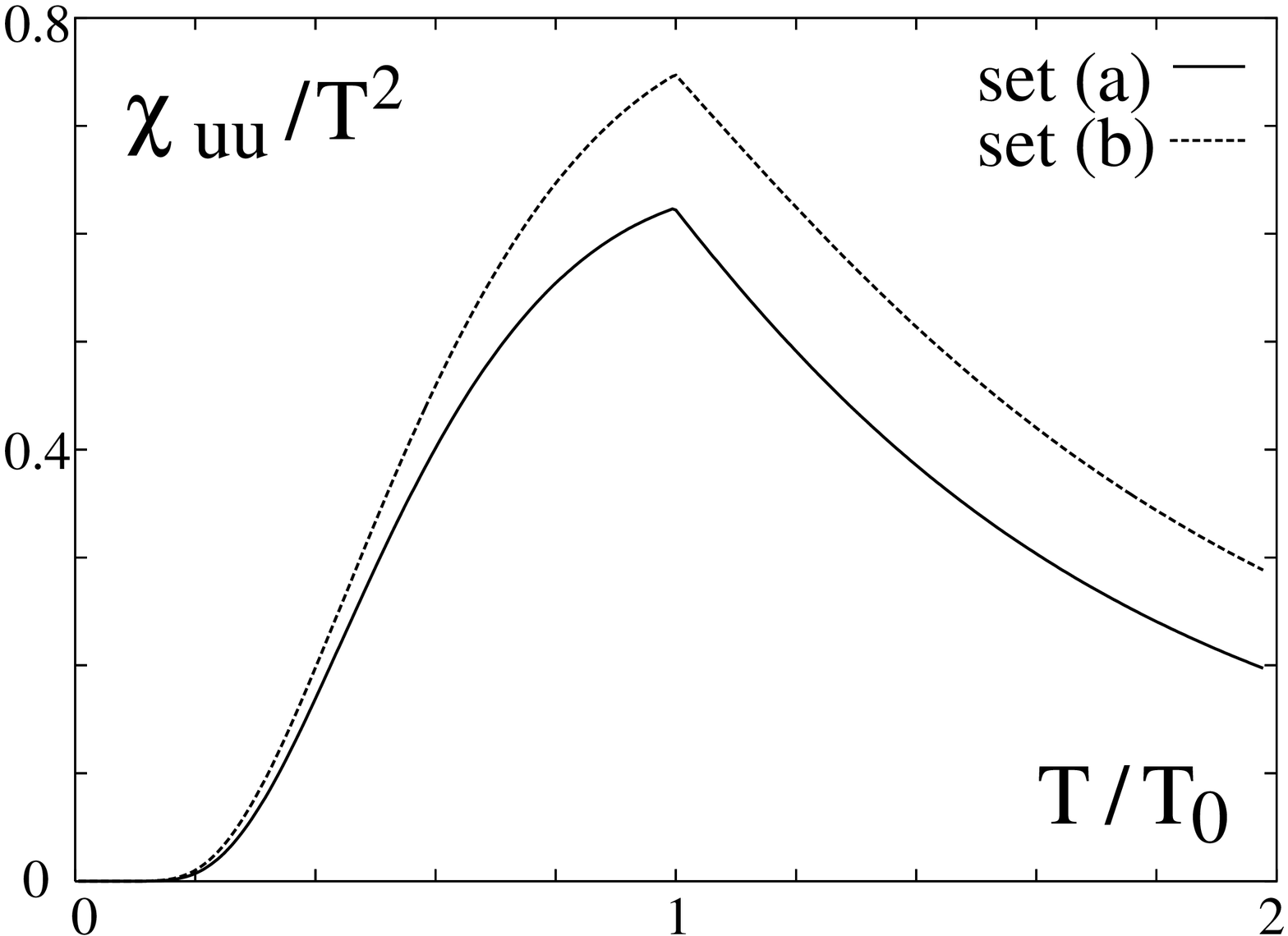}
\includegraphics[width=8cm]{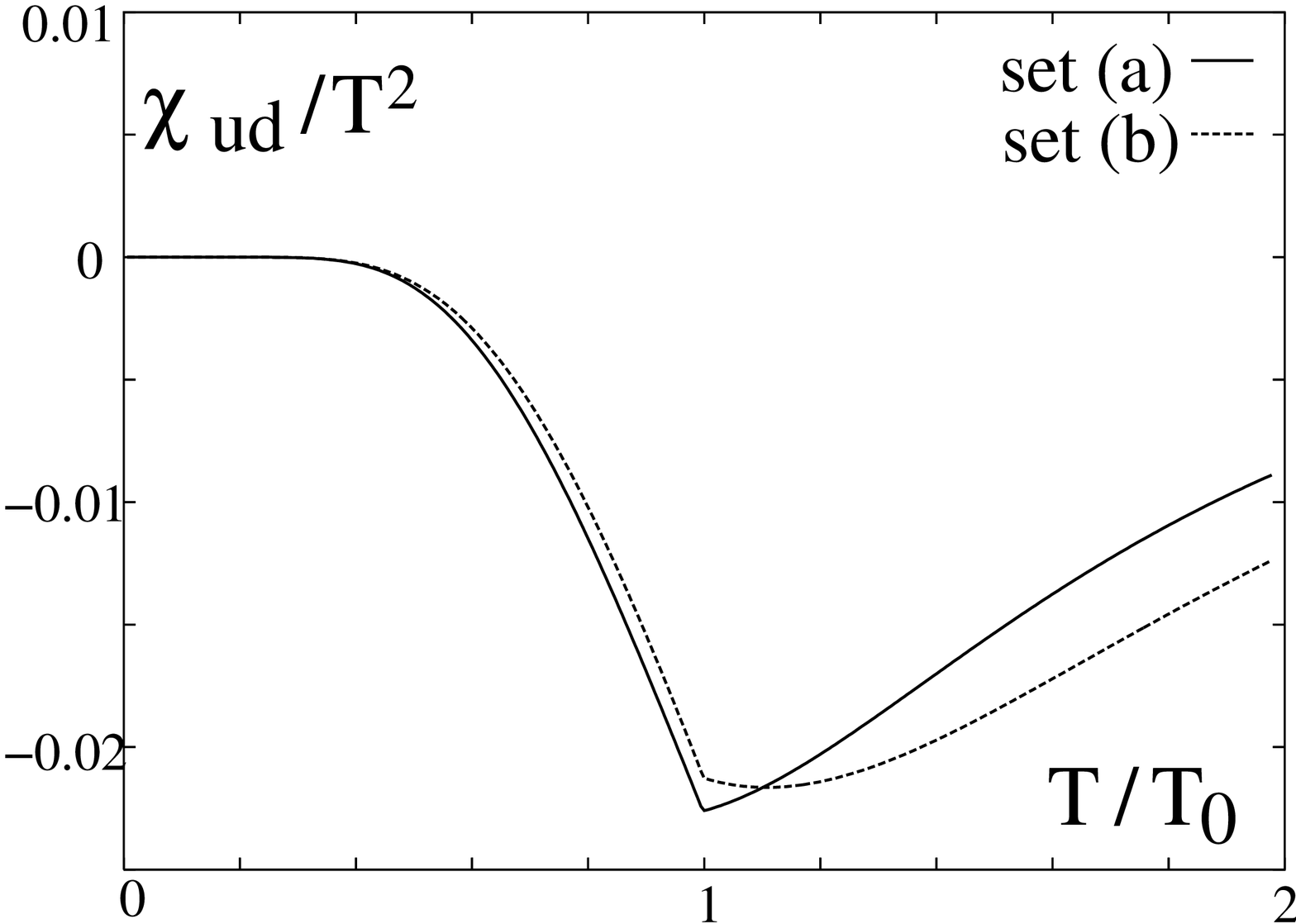}
\end{center}
\caption{\protect\label{fig:cutoff} The susceptibilities for
isospin symmetric system in the chiral limit  normalized to
$\chi^{\rm (free)}$ at $\mu_q=0$ as a function of $T/T_0$   for
$G_V^{\rm (V)}/G_V^{\rm (S)}=1/3$. The solid line corresponds to
the parameter set (a) and the dashed line to the set (b) listed in
Table~\ref{table:cutoff}. } 
\end{figure}

At vanishing chemical potential there are  generic features in the
temperature dependence of the quark flavor susceptibilities.
Consider the net quark $\chi_q$, the isovector $\chi_I$ and flavor
diagonal $\chi_{uu}$ susceptibilities shown in Figs.
\ref{fig:chiq}, \ref{fig:chiI} and \ref{fig:uu-muq}. Clearly all
these susceptibilities are strongly enhanced near the chiral phase
transition point $T_0$, while above $T_0$ the fluctuations are
suppressed. The increase of the quark susceptibilities with
temperature seen in the broken symmetry phase reflects a decrease
of the dynamical quark mass as the chiral transition is approached.
Consequently, the enhancement of the quark fluctuations is, in this
model, mainly due to the amplification of the Boltzmann factor
$\exp(-M/T)$ as the constituent quark mass is reduced. The
suppression of $\chi_q$, $\chi_I$ and $\chi_{uu}$ above $T_c$ is
due to the finite momentum cut-off of the NJL model. Indeed, the
flavor-diagonal susceptibility $\chi_{uu}$ in an ideal massless
quark gas with momentum cut--off $\Lambda$ is given by
\begin{eqnarray}
\chi_{uu}^{\rm (free)}/T^2 &=& \frac{12}{T^3}\int_0^\Lambda \frac{d^3
p}{(2\pi)^3} \frac{e^{p/T}}{\left( 1 + e^{p/T}
\right)^2}\nonumber\\
&\simeq&1-\frac{12}{\pi^2}e^{-\Lambda/T}(1+\frac {\Lambda}{T}+\frac
12\frac{\Lambda^2}{T^2})+\dots\,,
\label{free}
\end{eqnarray}
where the ellipsis in the second stands for terms that are
exponentially suppressed for $T<\Lambda$. For $\Lambda\to
\infty$ one obtains the ideal gas result $\chi_{uu}^{\rm
(free)}/T^2 = 1$, while for finite $\Lambda$ the fluctuation are
suppressed. At low temperature $T \ll
\Lambda$, the correction terms in Eq.~(\ref{free}) are negligible,
and $\chi$ is independent of $\Lambda$. However, for temperatures
on the order of $\Lambda$ there is a strong dependence on the
cut--off. With increasing temperature the suppression of $\chi_q$,
$\chi_I$ and $\chi_{uu}$ grows stronger, as seen in
Fig.~\ref{fig:cutoff}. As seen in Fig.~\ref{fig:cutoff} this holds
also for $\chi_q$ and  $\chi_I$. However, the cutoff dependence of
$\chi_{ud}$  is different. Here, since $\chi_{ud}\propto
\chi_q-\chi_I$, the cut--off dependence of $\chi_q$ and
$\chi_I$ partly cancel. Thus, the trend above and below $T_0$ is
different. At low temperatures the magnitude of $\chi_{ud}$ is
reduced and at high temperatures enhanced with increasing
$\Lambda$. We remind the reader that both parameter sets (a) and
(b) yield the same transition temperature $T_0
= 177$ MeV at $\mu_q = 0$ (see Table 1).

In Figs.~\ref{fig:uu-muq} and \ref{fig:ud-muq} we also show the
dependence of $\chi_{uu}$ and $\chi_{ud}$ on the choice  of vector
couplings. The dependence of $\chi_{uu}$ on $G_V^{\rm (V)}$ is
fairly weak, as seen in Fig.~\ref{fig:uu-muq}. This is because in
isospin symmetric matter ($\mu_I=0$) $G_V^{\rm (V)}$ contributes
only to $\chi_I$, which is much smaller than $\chi_q$ (cf. Fig.
\ref{tcp_q}). Consequently, $\chi_{uu}$ is essentially determined
by $\chi_q$, which depends on $G_S$ and $G_V^{\rm (S)}$.

A comparison of Fig.~\ref{fig:uu-muq} and \ref{fig:ud-muq} shows
that the  off-diagonal susceptibility $\chi_{ud}$ is much smaller
in magnitude than $\chi_{uu}$. Still, $\chi_{ud}$ is an interesting
observable, which may be used to identify the transition point.
This is particularly the case at finite $\mu_q$ where the
off-diagonal susceptibility is changing sign when crossing the
critical temperature. This behavior is consistent with the recent
LGT findings which shows negative values of $\chi_{ud}$ below and
above deconfinement for $\mu_q=0$. At finite $\mu_q$ and at $T=T_c$
the LGT results  show \cite{lattice:ejiri}  an abrupt change of
$\chi_{ud}$ from negative to positive value.  At the tricritical
point, $\chi_{ud}$, being proportional to $\chi_q -
\chi_I$, diverges as $\chi_q$  (see  Fig.~\ref{fig:ud-muq} (d)). Also
seen in Fig.~\ref{fig:ud-muq} is a rather strong variation of
$\chi_{ud}$ with the strength of $G_V^{\rm (S)}/G_V^{\rm (V)}$
ratio. Above the chiral phase transition, $\chi_q$ and $\chi_I$ are
equal for $G_V^{\rm (S)}=G_V^{\rm (V)}$. Hence, the fact that in
LGT $\chi_{ud}$ is very small above $T_0$, may be interpreted as a
signature of the universality of the isosclar and isovector
current-current interaction in the chirally restored phase. For
$\mu_q=0$, $\chi_{ud}$ vanishes also in the chirally broken phase
if $G_V^{\rm (S)}=G_V^{\rm (V)}$. The fact that in LGT $\chi_{ud}$
is negative in this temperature range, is consistent with $G_V^{\rm
(S)}>G_V^{\rm (V)}$, as expected for baryons from large $N_c$
arguments.

\subsection{  Susceptibilities at finite current quark mass}

So far, we have computed the susceptibilities in the chiral limit,
i.e., using an effective Lagrangian with an exact symmetry, the
chiral symmetry. However, the chiral symmetry of the QCD Lagrangian
is only approximate due to the finite current quark masses. In the
following, we account for the explicit chiral symmetry breaking and
explore its influence on the quark number fluctuations. This study
could be also interesting from the perspective of recent LGT
results \cite{lattice:ejiri} where the susceptibilities were
calculated for finite and large quark masses.

\begin{figure}
\begin{center}
\includegraphics[width=8cm]{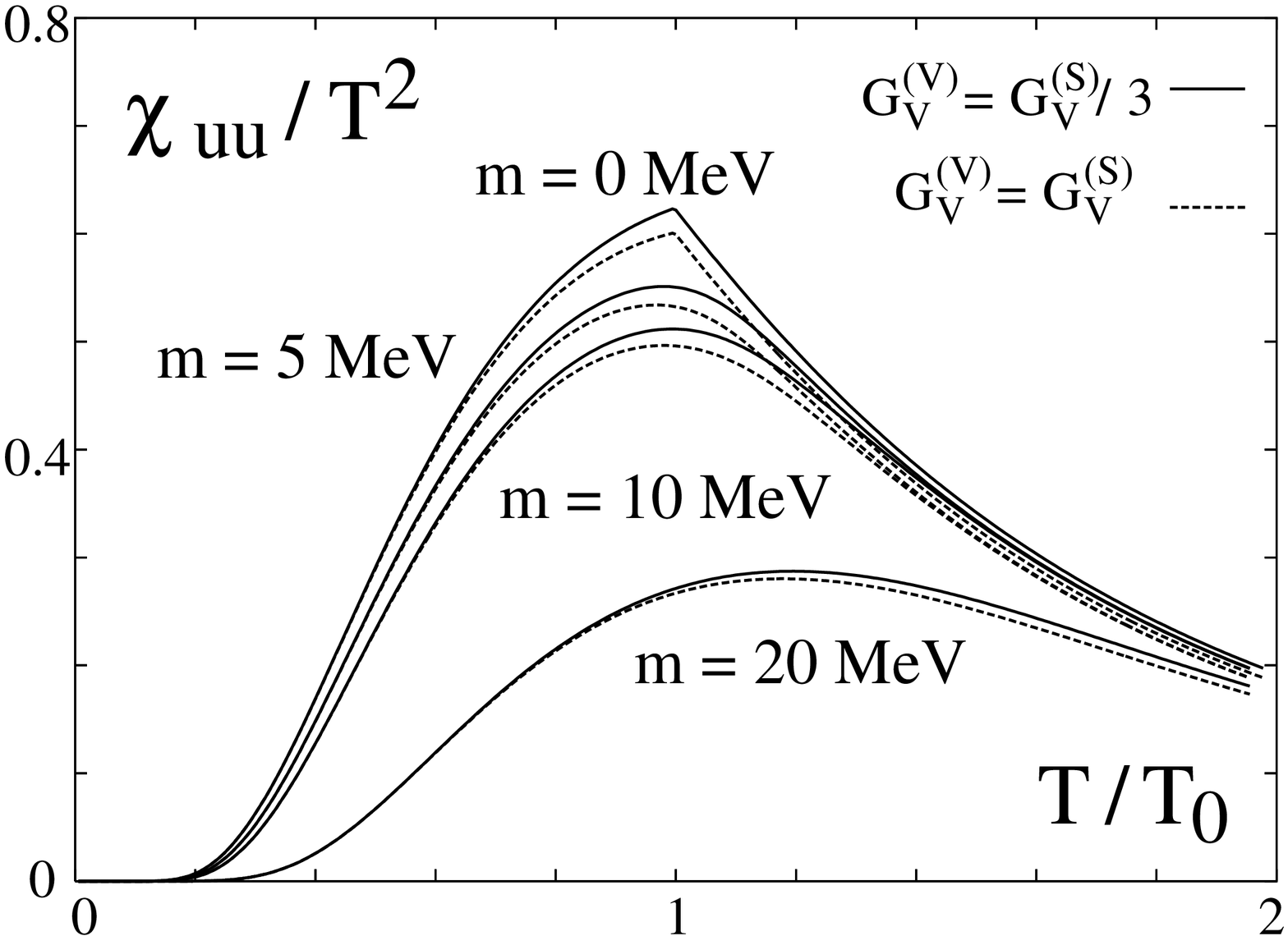}
\includegraphics[width=8cm]{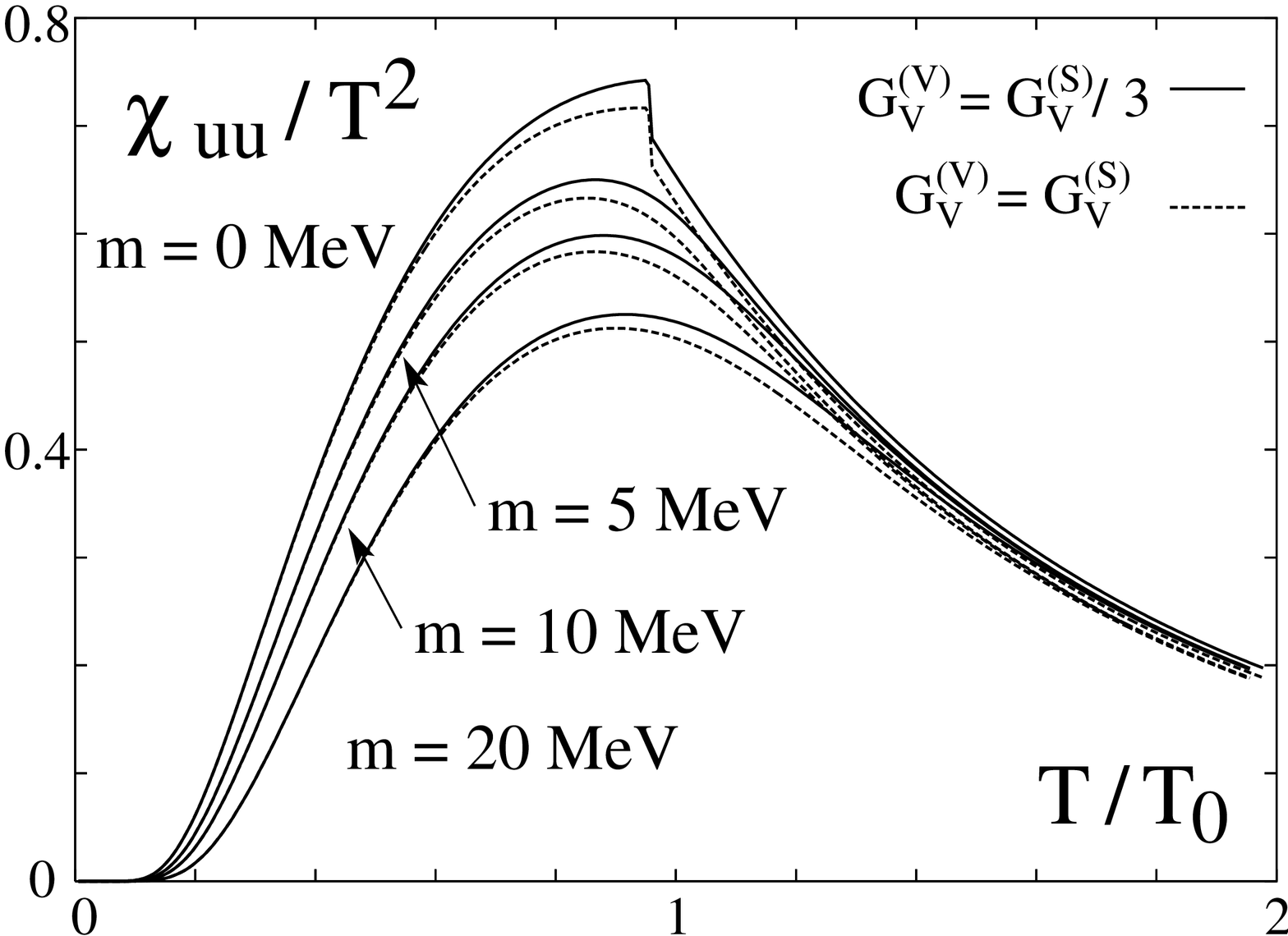}
\\
(a) $\mu_q = 0$ MeV
\hspace*{3.5cm}
(b) $\mu_q = 100$ MeV
\\
\includegraphics[width=8cm]{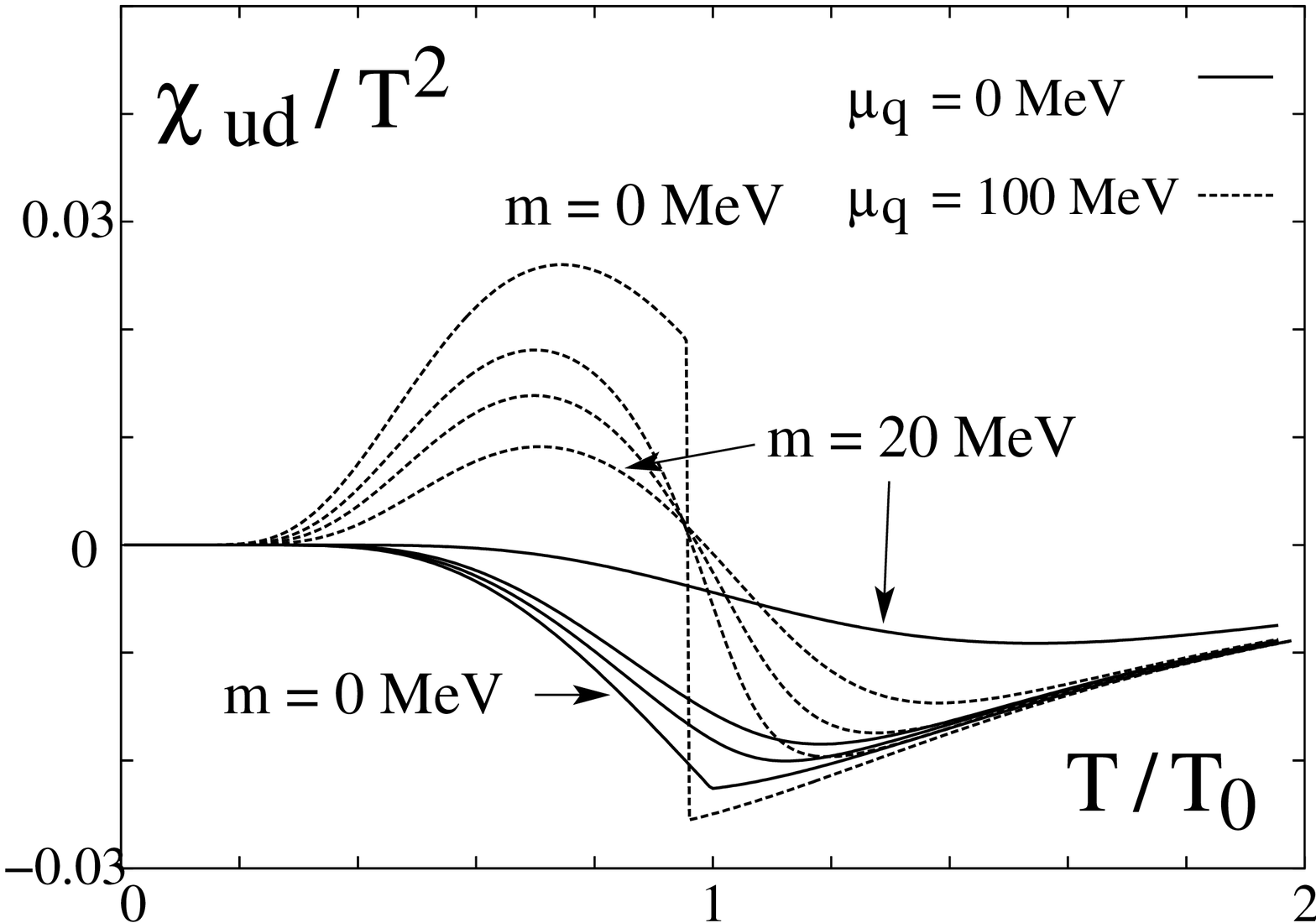}
\includegraphics[width=8cm]{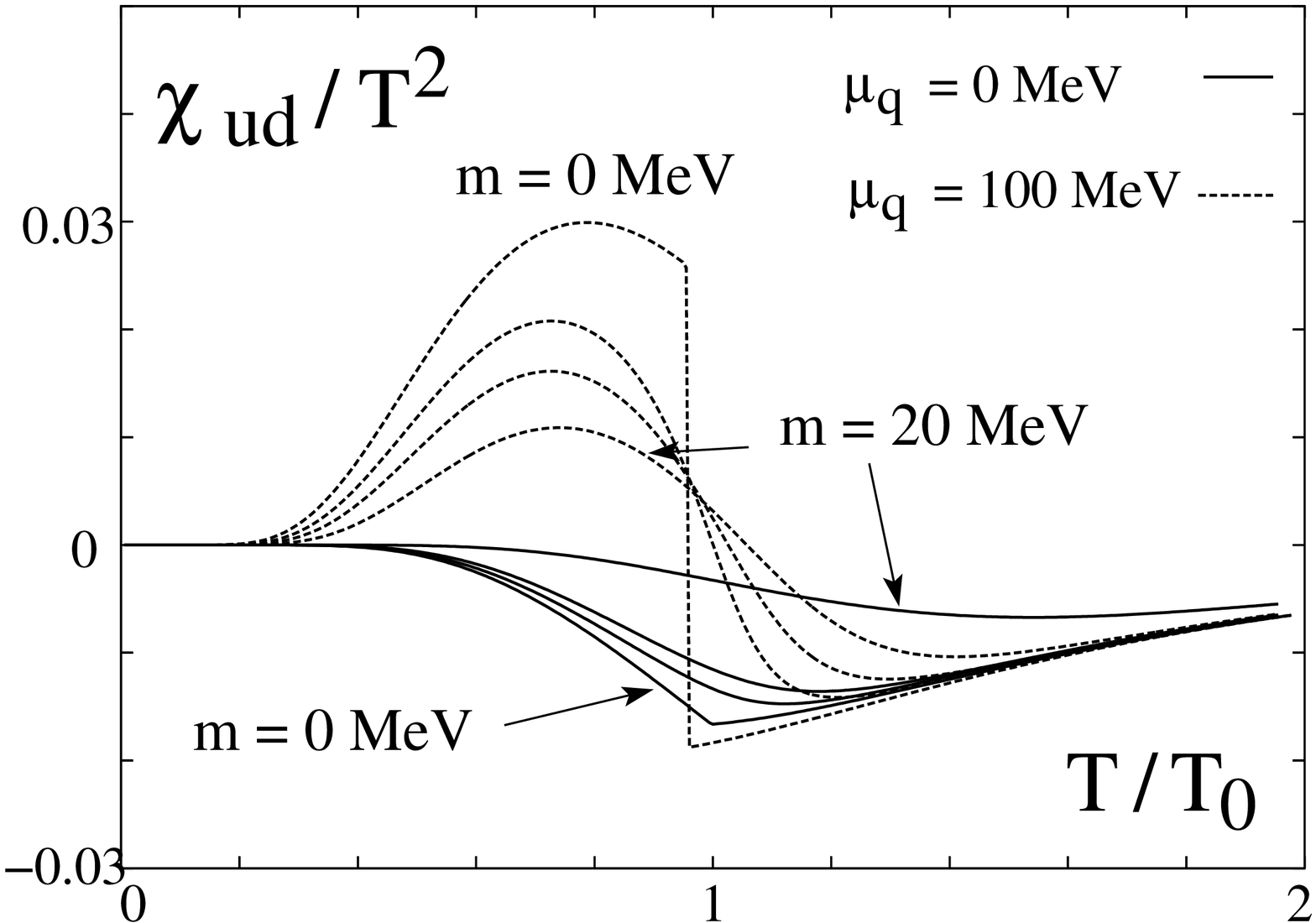}
\\
(c) $G_V^{\rm (V)}=G_V^{\rm (S)}/3$
\hspace*{3.5cm}
(d) $G_V^{\rm (V)}=G_V^{\rm (S)}/2$
\\
\includegraphics[width=8cm]{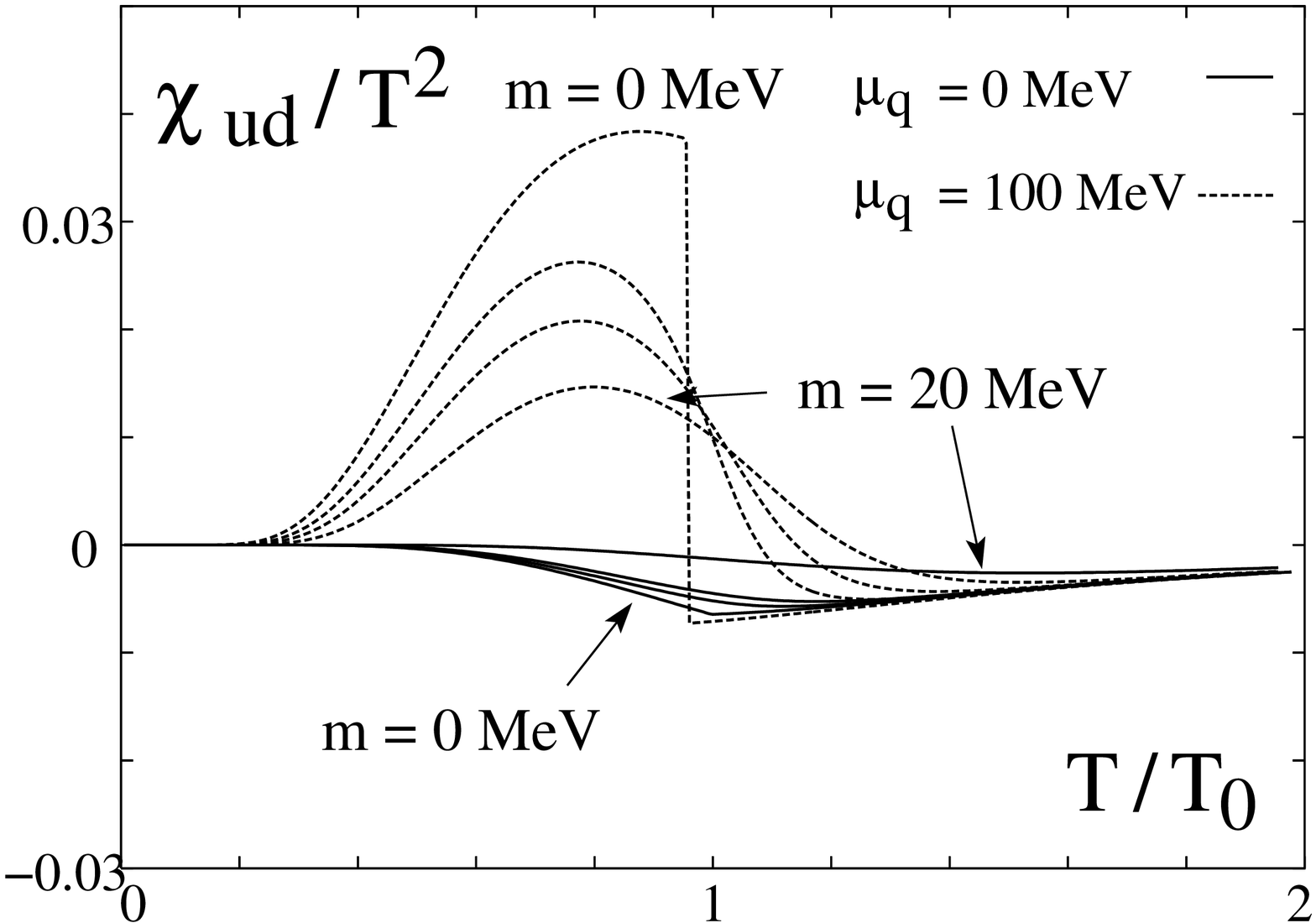}
\includegraphics[width=8cm]{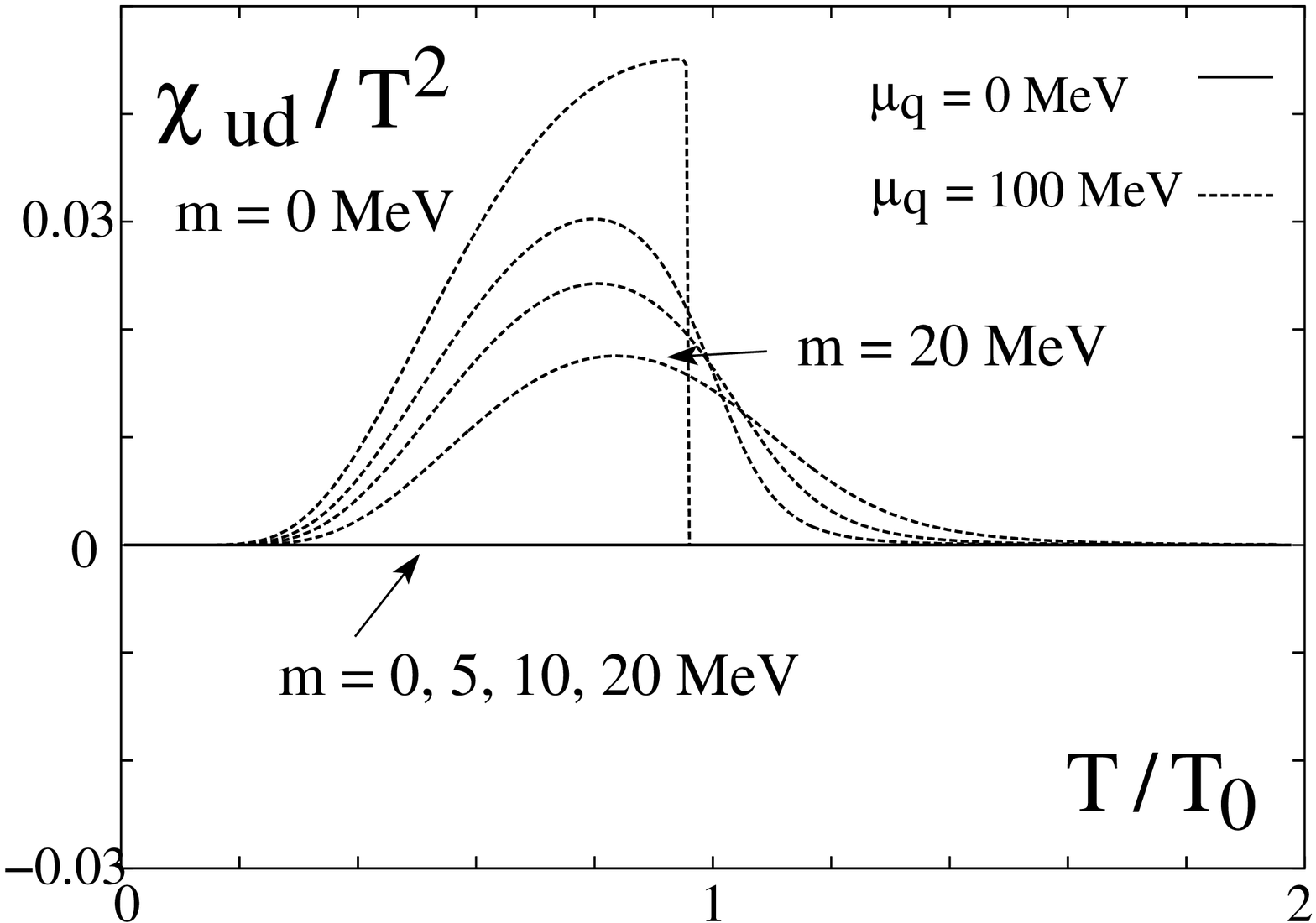}
\\
(e) $G_V^{\rm (V)}=G_V^{\rm (S)}/1.25$
\hspace*{3cm}
(f) $G_V^{\rm (V)}=G_V^{\rm (S)}$
\end{center}
\caption{\protect\label{fig:uu-m0} The  diagonal and off-diagonal
susceptibilities normalized to $T^2$ for different values of the
current quark mass as a function of $T/T_0$. For the
pseudo-critical temperature for $m = 5$ MeV we find $T_0
= 179$ MeV. The results correspond to $\mu_q = 0$ and $100$ MeV
and the vector coupling constants $G_V^{\rm (S)}=0.3\,G_S$ with
$G_V^{\rm (V)}/G_V^{\rm (S)}=1/3$ and $1$.
 }
\end{figure}

In Fig.~\ref{fig:uu-m0} we show results obtained with the NJL model
for the flavor diagonal $\chi_{uu}$ and off-diagonal $\chi_{ud}$
susceptibilities for different values of the current quark mass.
These calculations were done both for vanishing and finite chemical
potential as well as for several sets of the vector coupling
constant.

A comparison of the results for different quark masses clearly
shows that $\chi_{uu}$ exhibits a peak structure at finite
temperature, indicating the phase change, for all values of $m$.
However, at finite $m$ the $\chi_{uu}$ is a smooth function
everywhere, whereas in the chiral limit it exhibits a non-analytic
structure at the phase transition. This behavior indicates that
the second order transition at $m=0$ is converted into a smooth
cross-over at finite $m$. The above is even more transparent at
finite $\mu_q$ where the discontinuity of $\chi_{uu}$ at $T_c$
disappears at finite quark mass.  The behavior of the NJL model at
finite $m$ is in accord with recent LGT finding in 2-flavor QCD,
which also shows a cross-over transition at finite quark mass. We
note, however, that the second-order transition expected in the NJL
model in the chiral limit is still not confirmed by  LGT
calculations. Some recent results even suggest that in 2--flavor
QCD this transition could be weakly first order \cite{digiacomo}.

With increasing quark mass the peak position of $\chi_{uu}/T^2$ is
shifted towards larger $T$, both at vanishing and at finite
$\mu_q$. The height of the peak also depends on $m$ and decreases
with increasing current quark mass. The shift of the peak position
in $\chi_{uu}$ indicates an increase of the  pseudo-critical
temperature with increasing quark mass. Such an effect was observed
in lattice calculations already some time ago~\cite{paikert}. Also
the suppression of quark number fluctuations with increasing quark
mass is found in recent LGT calculations both at vanishing and at
finite $\mu_q$. This suppression is related to an upward shift of
the hadronic mass spectrum~\cite{rg2}, which leads to a reduction
of the number of thermally excited hadrons $\sim\exp(-M_h/T)$. The
suppression of $\chi_{uu}$ in the NJL model is of similar origin;
the number of thermally excited quarks is cut back by the
corresponding increase of the dynamical quark mass. The increase of
$\chi_{uu}$ with $\mu_q$ seen in Fig.~\ref{fig:uu-m0} can be
understood in terms of the corresponding increase of the thermal
factors $\sim\exp(\mu_q/T)$. Finally, at large temperatures $T>T_0$
the NJL model shows a very weak dependence of the susceptibilities
on the current quark mass independently of the value of $\mu_q$. In
lattice calculations of QCD the $m_q$--dependence of thermodynamic
quantities was also found to be weak for $m_q/T<1$.

The influence of a non-zero current quark mass on the off-diagonal
susceptibility $\chi_{ud}$  for different values of $\mu_q$ and
vector couplings is illustrated in Figs.~\ref{fig:uu-m0} (c)-(f).
For $\mu_q=0$ and $G_V^{\rm (V)}/G_V^{\rm (S)} < 1$, $\chi_{ud}$ is
finite and negative for all $T$, and approaches zero for large
temperatures. A similar behavior is observed in 2-flavor QCD on the
lattice \cite{lattice:ejiri}. When $G_V^{\rm (V)}/G_V^{\rm (S)}$ is
increased towards unity, $\chi_{ud}$ approaches zero at all
temperatures. At finite $\mu_q$, the temperature dependence of
$\chi_{ud}$ changes qualitatively. Below $T_c$ it is negative,
while above $T_c$ it is positive, in qualitative agreement with the
results of LGT calculations \cite{lattice:ejiri}. The off-diagonal
susceptibility $\chi_{ud}$ is finite near the pseudo-critical
transition for all values of $m$.


When discussing the influence of the model parameters on quark
number susceptibilities we have allowed for variations of the
parameters. However, we have not considered a possible $T$ and
$\vec\mu$ dependence of $G_S$,$G_V$ and $\Lambda$. It was recently
argued that such a dependence is important for a quantitative
comparison of NJL model results with LGT findings \cite{sh}.
However, so far systematic calculations of the variation of these
parameters with temperature and chemical potential are lacking. A
possible temperature dependence of these parameters was obtained
phenomenologically by comparing some observables with lattice
results.


\section{Summary and Conclusions}
\label{sec:SD}

We  have discussed the properties of quark number fluctuations
within the framework of the Nambu--Jona-Lasinio (NJL) model. The
model was formulated at finite temperature and chemical potentials
for baryon number and isospin. In a mean field approach, we have
shown how the fluctuations of different quark flavors are changing
across the phase boundary. Such a study is of interest from the
perspective of both heavy ion phenomenology and lattice gauge
theory. In the first case we have explored the non-monotonic
structure of net quark, diagonal and off-diagonal susceptibilities
along the phase transition line. We have also discussed the
critical region around the tricritical point in the context of
heavy ion phenomenology.

The results on different quark susceptibilities at finite quark
mass are in qualitative agreement with recent findings on the
lattice. Our study may give some insight into how lattice results
may change when one approaches the chiral limit at vanishing and at
large baryon chemical potential. This expectation is supported by
the fact that the NJL model exhibits the same critical properties
as one expects for QCD.

Obviously, the NJL model differs substantially from QCD. This model
does not contain all relevant hadronic degrees of freedom, which in
QCD contribute substantially to the quark number susceptibilities.
Moreover, the phase-space of dynamical quarks is suppressed by the
ultraviolet cut--off. Consequently, the perturbative regime of QCD
at high temperature and density is not reproduced by this model.
Nevertheless, features of the quark number susceptibilities that
probe the restoration of chiral symmetry can be studied in some
detail in NJL model calculations. In particular the change in the
behavior of the quark fluctuations near the critical point can be
interpreted as an effective change of the vector interaction
associated with the chiral phase transition.


\section*{ Acknowledgments}

We acknowledge interesting discussions with F. Karsch, S. Leupold,
and J. Wambach. C.S. also acknowledges fruitful   discussions with H.
Fujii and B.~J.~Schaefer. The work of B.F. and C.S. were supported
in part by the Virtual Institute of the Helmholtz Association
under the grant No. VH-VI-041. K.R. acknowledges partial support
of the Gesellschaft f\"ur Schwerionenforschung (GSI)   and KBN
under grant 2P03 (06925).


\appendix

\setcounter{section}{0}
\renewcommand{\thesection}{\Alph{section}}
\setcounter{equation}{0}
\renewcommand{\theequation}{\Alph{section}.\arabic{equation}}

\section{Analytical results for the phase boundary}
\label{app:phase-boundary}

For massless quarks, the gap equation (\ref{gap eq-mass}) at the
chiral transition can be obtained in closed form. At the second
order transition, the gap equation has a non-trivial solution at
$M=0$, i.e.
\begin{equation}
1=4 G_S N_c \sum_f\int\frac{d^3p}{(2\pi)^3}\frac 1p
\Bigl[1-n^{(+)}_f(\vec{p},T,\tilde{\mu}_f)-n^{(-)}_f(\vec{p},T,\tilde{\mu}_f)\Bigl]
\end{equation}
is satisfied. After some rearrangement one finds for
$\tilde{\mu}_u=\tilde{\mu}_d=\tilde{\mu}$
\begin{equation}
\frac{\Lambda^2}{2}-\frac{\pi^2}{4 G_S N_c}=I(T,\tilde{\mu})\, ,
\label{reduced-gap}
\end{equation}
where the quadratically divergent term on the left hand side is due
to the vacuum loop and $I(T,\tilde{\mu})$ to the thermal loops. The
explicit form of the latter is given by
\begin{align}
I(T,\mu)&=\int_0^\Lambda dp
p\left(\frac{1}{e^{\beta(p-\mu)}+1}+\frac{1}{e^{\beta(p+\mu)}+1}\right)\nonumber\\
&=\frac{\pi^2 T^2}{6} +\frac{\mu^2}{2} +
T^2\left[\mbox{L}_2(-e^{-\beta(\mu+\Lambda)})+\mbox{L}_2(-e^{\beta(\mu-\Lambda)})\right]\nonumber\\
&-T\Lambda\log\left[(1+e^{-\beta(\mu+\Lambda)})(1+e^{\beta(\mu-\Lambda)})\right]\,
,
\label{closed-gap}
\end{align}
where $\mbox{L}_2[z]=\sum_{n=1}^\infty z^n/n^2$ is Euler's
dilogarithm \cite{Erdelyi}.

We now use (\ref{closed-gap}) to explore the dependence of the
phase boundary on the cut off $\Lambda$. A general variation of
(\ref{reduced-gap}) yields
\begin{equation}
\Lambda\delta\Lambda +\frac{\pi^2}{4G_S^2N_c}\delta G_s = \frac{\partial I}{\partial \Lambda}\delta \Lambda+
\frac{\partial I}{\partial T}\delta T +\frac{\partial I}{\partial \mu}\delta \mu\, .
\end{equation}
At $\mu=0$, $\partial I/\partial\mu = 0$ by symmetry and
\begin{equation}
\frac{\partial I}{\partial\Lambda}=\frac{2}{e^{\beta\Lambda}+1}\,\Lambda\, .
\end{equation}
Thus, the requirement that the critical temperature at vanishing
net quark density remains fixed, as in the right panel of Fig.
\ref{fig:phase}, implies the following  relation between the cut
off $\Lambda$ and the coupling constant $G_S$
\begin{equation}
\Lambda\delta\Lambda +\frac{\pi^2}{4G_S^2N_c}\delta G_s =\frac{2}{e^{\beta\Lambda}+1}\Lambda\delta\Lambda\, .
\end{equation}
Assuming that the transition is second order everywhere, the shift
of the critical value of the chemical potential at $T=0$ is given
by
\begin{equation}
\delta\mu=\frac 1\mu\left(\Lambda\delta\Lambda +\frac{\pi^2}{4G_S^2N_c}\delta G_s\right) =
\frac{2}{e^{\Lambda/T}+1}\frac{\Lambda}{\mu}\,\delta\Lambda\, .
\end{equation}
This relation, which remains approximately valid also when the
transition is weakly first order, provides a quantitative
interpretation of the shift of the phase boundary shown in the
right panel of Fig. \ref{fig:phase}.

Using the fact that for reasonable parameter choices,
$e^{\beta(\pm\mu-\Lambda)}\ll 1$ along the phase boundary, useful
approximative expressions may be obtained for $I(T,\mu)$. Retaining
the first two terms in the expansions of the dilogarithms and the
logarithm in (\ref{closed-gap}), we find
\begin{align}
I(T,\mu)&=\frac{\pi^2 T^2}{6} +\frac{\mu^2}{2}
-T(T+\Lambda)\left(e^{\beta(\mu-\Lambda)}+e^{-\beta(\mu+\Lambda)}\right)\nonumber\\
&+\frac 14
T(T+2\Lambda)\left(e^{2\beta(\mu-\Lambda)}+e^{-2\beta(\mu+\Lambda)}\right)\,
.
\label{th-loop-appr}
\end{align}
Within this approximation, one finds the critical temperature at
$\mu=0$ with an accuracy of $\sim 10^{-5}$. For non-zero $\mu$ the
approximation becomes increasingly better, since the
$\Lambda$-dependent terms drop out for $T\rightarrow 0$. We note in
passing that a solution of Eqn. (\ref{reduced-gap}) is possible
only if the left hand side is positive, i.e., for $G_S
\Lambda^2 >\pi^2/(2 N_c)\simeq 1.64$. For smaller values of the
scalar coupling constant, chiral symmetry is never broken.



\section{Derivatives of effective condensates}
\label{app:der}

In this appendix, we summarized  derivatives of the dynamical
masses $M_f$ and the shifted chemical potentials $\tilde{\mu}_f$.
These results are  obtained  from  the gap equations~(\ref{gap
eq-mass})- (\ref{muI}) by taking derivatives with respect to
$\mu_q$ and $\mu_I$ as
\begin{align}
&\frac{\partial M_f}{\partial\mu_{q,I}}
= \frac{\partial M_f}{\partial\mu_{q,I}}A_f
{}+ \frac{\partial M_{f^\prime}}{\partial\mu_{q,I}}A_{f^\prime}
{}+ \frac{\partial\tilde{\mu}_f}{\partial\mu_{q,I}}B_f
{}+ \frac{\partial\tilde{\mu}_{f^\prime}}{\partial\mu_{q,I}}
    B_{f^\prime}\,,
\nonumber\\
&1
= \frac{\partial\tilde{\mu}_u}{\partial\mu_{q,I}}
{}+ \frac{\partial M_u}{\partial\mu_{q,I}}C_u^{\rm (S)}
{}+ \frac{\partial M_u}{\partial\mu_{q,I}}C_u^{\rm (V)}
{}+ \frac{\partial M_d}{\partial\mu_{q,I}}C_d^{\rm (S)}
{}- \frac{\partial M_d}{\partial\mu_{q,I}}C_d^{\rm (V)}
\nonumber\\
&\qquad
{}+ \frac{\partial\tilde{\mu}_u}{\partial\mu_{q,I}}D_u^{\rm (S)}
{}+ \frac{\partial\tilde{\mu}_u}{\partial\mu_{q,I}}D_u^{\rm (V)}
{}+ \frac{\partial\tilde{\mu}_d}{\partial\mu_{q,I}}D_d^{\rm (S)}
{}- \frac{\partial\tilde{\mu}_d}{\partial\mu_{q,I}}D_d^{\rm (V)}\,,
\nonumber\\
&\pm 1
= \frac{\partial\tilde{\mu}_d}{\partial\mu_{q,I}}
{}+ \frac{\partial M_u}{\partial\mu_{q,I}}C_u^{\rm (S)}
{}- \frac{\partial M_u}{\partial\mu_{q,I}}C_u^{\rm (V)}
{}+ \frac{\partial M_d}{\partial\mu_{q,I}}C_d^{\rm (S)}
{}+ \frac{\partial M_d}{\partial\mu_{q,I}}C_d^{\rm (V)}
\nonumber\\
&\qquad
{}+ \frac{\partial\tilde{\mu}_u}{\partial\mu_{q,I}}D_u^{\rm (S)}
{}- \frac{\partial\tilde{\mu}_u}{\partial\mu_{q,I}}D_u^{\rm (V)}
{}+ \frac{\partial\tilde{\mu}_d}{\partial\mu_{q,I}}D_d^{\rm (S)}
{}+ \frac{\partial\tilde{\mu}_d}{\partial\mu_{q,I}}D_d^{\rm (V)}\,,
\label{diff-eq}
\end{align}
where $f \neq f^\prime \in \{u,d\}$ and
the functions $A_f, B_f, C_f^{\rm (S,V)}$ and $D_f^{\rm (S,V)}$
are defined by
\begin{align}
&A_f
= 4G_S N_c \int\frac{d^3p}{(2\pi)^3} \frac{1}{E_f}
\Biggl[\Biggl( 1 - \frac{M_f^2}{E_f^2} \Biggr)
 \bigl( 1 - n_f^{(+)} - n_f^{(-)} \bigr)
\nonumber\\
&\qquad\qquad
 {}+ \frac{M_f^2}{T E_f} \Bigl[ n_f^{(+)}\bigl( 1 - n_f^{(+)} \bigr)
 {}+   n_f^{(-)}\bigl( 1 - n_f^{(-)} \bigr)
\Bigr] \Biggr]\,,
\nonumber\\
&B_f
= 4G_S N_c \int\frac{d^3p}{(2\pi)^3}\frac{-M_f}{T E_f}
\Biggl[ n_f^{(+)}\bigl( 1 - n_f^{(+)} \bigr)
{}- n_f^{(-)}\bigl( 1 - n_f^{(-)} \bigr) \Biggr]\,,
\nonumber\\
&C_f^{\rm (S,V)}
= 4G_V^{\rm (S,V)} N_c \int\frac{d^3p}{(2\pi)^3}\frac{-M_f}{T E_f}
\Biggl[ n_f^{(+)}\bigl( 1 - n_f^{(+)} \bigr)
{}- n_f^{(-)}\bigl( 1 - n_f^{(-)} \bigr) \Biggr]\,,
\nonumber\\
&D_f^{\rm (S,V)}
= 4G_V^{\rm (S,V)} N_c \int\frac{d^3p}{(2\pi)^3}\frac{1}{T}
\Biggl[ n_f^{(+)}\bigl( 1 - n_f^{(+)} \bigr)
{}+ n_f^{(-)}\bigl( 1 - n_f^{(-)} \bigr) \Biggr]\,.
\label{func}
\end{align}
Solving Eq.~(\ref{diff-eq}) one gets,
\begin{align}
\frac{\partial M_u}{\partial\mu_{q,I}}
&= \Bigl[ B_u (1 + 2D_d^{\rm (V,S)})
{}\pm B_d (1 + 2D_u^{\rm (V,S)}) \Bigr]/F\,,
\nonumber\\
\frac{\partial M_d}{\partial\mu_{q,I}}
&= \frac{\partial M_u}{\partial\mu_{q,I}}
\nonumber\\
\frac{\partial\tilde{\mu}_u}{\partial\mu_{q,I}}
&= \Bigl[ 1 - (A_u + A_d)(1 + 2D_d^{\rm (V,S)})
{}\mp 2B_d(C_u^{\rm (V,S)}
{}\mp C_d^{\rm (V,S)}) + 2D_d^{\rm (V,S)} \Bigr]/F\,,
\nonumber\\
\frac{\partial\tilde{\mu}_d}{\partial\mu_{q,I}}
&= \pm \Bigl[ 1 - (A_u + A_d)(1 + 2D_u^{\rm (V,S)})
{}+ 2B_u(C_u^{\rm (V,S)} \mp C_d^{\rm (V,S)})
{}+ 2D_u^{\rm (V,S)} \Bigr]/F\,,
\label{diff}
\end{align}
with the function $F$ expressed by
\begin{align}
F
&= 1 - (A_u + A_d)\left[ 1 + D_u^{\rm (S)}(1 + 2D_d^{\rm (V)})
{}+ D_u^{\rm (V)}(1 + 2D_d^{\rm (S)}) + D_d^{\rm (S)}
{}+ D_d^{\rm (V)} \right]
\nonumber\\
&
{}+ B_u \left[ C_u^{\rm (S)} + C_d^{\rm (S)} + C_u^{\rm (V)}
{}- C_d^{\rm (V)}\right]
{}+ B_d \left[ C_u^{\rm (S)} + C_d^{\rm (S)} - C_u^{\rm (V)}
{}+ C_d^{\rm (V)}\right]
\nonumber\\
&
{}+ 2\left[B_u D_d^{\rm (S)} - B_d D_u^{\rm (S)}\right]
    \left[C_u^{\rm (V)} - C_d^{\rm (V)}\right]
{}+ 2\left[B_u D_d^{\rm (V)} + B_d D_u^{\rm (V)}\right]
    \left[C_u^{\rm (S)} + C_d^{\rm (S)}\right]
\nonumber\\
&
{}+ D_u^{\rm (S)}\left[1 + 2D_d^{\rm (V)}\right]
{}+ D_d^{\rm (S)}\left[1 + 2D_u^{\rm (V)}\right]
{}+ D_u^{\rm (V)} + D_d^{\rm (V)}\,.
\label{denominator}
\end{align}
The  $\chi_q$ and $\chi_I$ susceptibilities are obtained by
substituting Eq.~(\ref{diff}) into Eqs.~(\ref{sus_q}) and
(\ref{sus_I}).

\section{Critical exponents in Landau theory}
\label{app:crit}
In this appendix we compute the mean-field critical exponents of
the TCP in Landau theory. Since the fourth order term $b(T,\mu_q)$
in (\ref{eqgl}) vanishes at the TCP, it is necessary to include a
sixth-order term in the expansion of the thermodynamic potential
\begin{equation}
\omega^{}(T,\mu_q,M) \simeq  \omega_0(T,\mu_q) {}+
\frac{1}{2}a(T,\mu_q)M^2 {}+
\frac{1}{4}b(T,\mu_q)M^4 {}+
\frac{1}{6}c M^6\,,\label{landau-tcp}
\end{equation}
where $M$ is the dynamical quark mass. We assume that $c>0$ and
neglect its dependence on the temperature and chemical potential.
At the TCP the coefficients $a$ and $b$ both vanish. Close to the
TCP we retain the leading terms,
\begin{align}
a(T,\mu)&=A(T-T_{TCP})+B(\mu-\mu_{TCP})\\\nonumber
b(T,\mu)&=C(T-T_{TCP})+D(\mu-\mu_{TCP})\, .
\end{align}
Along the second order phase boundary, the coefficient $a$
vanishes, while at the first order transition $a>0$ and $b<0$. The
dynamical mass is determined by minimizing the thermodynamic
potential. The solutions of the gap equation
\begin{equation}
\frac{\partial \omega}{\partial M} = M(a+b M^2+c M^4)=0
\end{equation}
are given by $M=0$ and
$M^2=-\frac{b}{2c}\pm\frac{1}{2c}\sqrt{b^2-4ac}$. The quark number
susceptibility is given by
\begin{equation}
\chi_q=-\frac{\partial^2 \omega}{\partial \mu_q^2} = \chi_0-\frac 12\left( B
+ D M^2\right)\frac{\partial M^2}{\partial \mu_q}\, .
\end{equation}
As the TCP is approached, $a,b$ and $M\rightarrow 0$. Thus, the
singular part of the susceptibility is given by
\begin{equation}
\label{appC:chi-sing}
\chi_q^{sing}= \frac 12 \frac{B^2}{\sqrt{b^2-4ac}}\, .
\end{equation}
When the TCP is approached along the second order phase boundary,
$a=0$. Consequently $\chi_q^{sing}= B^2/2|b|\sim
|\mu_q-\mu_{TCP}|^{-1}$. In order to compute the critical exponent
along the first order transition, we must first determine the
relation between $a$, $b$ and $c$ along this path.  At the first
order transition there are three degenerate minima of the
thermodynamic potential, one at $M=0$ corresponding to the
symmetric phase, and two at finite M corresponding to the two
realizations of the broken phase. Thus, we are looking for a
solution to the two equations
\begin{align}
&aM^2 {}+
\frac{1}{2}bM^4 {}+
\frac{1}{3}c M^6=0\\\nonumber
&a+b M^2+c M^4=0\, ,
\end{align}
for non-vanishing $M$. Such a solution, $M^2=-3b/4c$, exists when
the coefficients satisfy the relation $16 a c = 3 b^2$. This
relation defines the location of the phase boundary in the
$(T,\mu_q)$ plane. Furthermore, it implies that the first-order
phase boundary is asymptotically parallel to the second order one.
Using the relation in (\ref{appC:chi-sing}), we find
$\chi_q^{sing}=B^2/|b|\sim |\mu_q-\mu_{TCP}|^{-1}$. Thus, the
critical exponent is identical to that along the second order line,
but the pre-factor is twice as large. If the TCP is approached from
the broken phase, along any path which is not asymptotically
tangential to the phase boundary at the TCP, $b^2\ll 4 a c$ for
points close to the TCP. This implies that the critical exponent is
different, namely $\chi_q^{sing}
= B^{2}/4\sqrt{ac}\sim |\mu_q-\mu_{TCP}|^{-\frac 12}$.

Finally, we explore the scaling behavior of the dynamical quark
mass near the O(4) critical line and at the tricritical point. As
discussed in the text, this behavior determines the critical
properties of the quark number susceptibility. The non-trivial
solution of the gap equation implies that for constant chemical
potential $M^2\sim a\sim |T_c-T|$ at the O(4) critical line, where
$b\neq 0$. On the other hand, at the critical end point, where also
$b\rightarrow 0$, $M^2\sim \sqrt{a}\sim |T_{TCP}-T|^{\frac 12}$.


\end{document}